\documentclass{aastex62}

\usepackage{etoolbox}
\usepackage{graphicx}	
\usepackage{multirow}
\usepackage{cleveref}
\usepackage{CJK}

\newcommand{\cloudy}{\textsc{cloudy}}
\newcommand{\kms}{km s$^{-1}$}
\newcommand{\cmN}{cm$^{-2}$}

\newcommand{\lam}{$\lambda$}

\newcommand{\lya}{\mbox{Ly$\alpha$}}
\newcommand{\lyb}{\mbox{Ly$\beta$}}
\newcommand{\lyc}{\mbox{Ly$\gamma$}}
\newcommand{\lyd}{\mbox{Ly$\delta$}}
\newcommand{\lye}{\mbox{Ly$\epsilon$}}
\newcommand{\hi}{\mbox{H\,{\sc i}}}

\newcommand{\hb}{H$\beta$}

\newcommand{\civ}{\mbox{C\,{\sc iv}}}
\newcommand{\ciii}{\mbox{C\,{\sc iii}}}
\newcommand{\cii}{\mbox{C\,{\sc ii}}}
\newcommand{\siiv}{\mbox{Si\,{\sc iv}}}
\newcommand{\svi}{\mbox{S\,{\sc vi}}}
\newcommand{\siv}{\mbox{S\,{\sc iv}}}
\newcommand{\siiii}{\mbox{Si\,{\sc iii}}}
\newcommand{\siii}{\mbox{Si\,{\sc ii}}}
\newcommand{\nv}{\mbox{N\,{\sc v}}}

\newcommand{\niii}{\mbox{N\,{\sc iii}}}

\newcommand{\ovi}{\mbox{O\,{\sc vi}}}

\newcommand{\oiv}{\mbox{O\,{\sc iv}}}
\newcommand{\oiii}{\mbox{O\,{\sc iii}}}

\newcommand{\feii}{\mbox{Fe\,{\sc ii}}}

\newcommand{\mgi}{\mbox{Mg\,{\sc i}}}
\newcommand{\mgii}{\mbox{Mg\,{\sc ii}}}

\newcommand{\neviii}{\mbox{Ne\,{\sc viii}}}

\received{2019 April 11}
\revised{2019 October 14}
\accepted{2019 October 30}
\published{2019 December 12}

\submitjournal{ApJ}

\shorttitle{Nature and Origins of Rich \civ\ Complexes}
\shortauthors{C. Chen, F. Hamann, L. Simon and B. Ma}

\begin{document}


\title{Nature and Origins of Rich Complexes of \civ\ Associated Absorption Lines}

\correspondingauthor{Chen Chen}
\email{chenchen8@mail.sysu.edu.cn}
\email{mabo8@mail.sysu.edu.cn}

\author{Chen Chen}
\affiliation{School of Physics \&\ Astronomy\\
Sun Yat-Sen University\\ 
ZhuHai 519000, China}

\affiliation{Department of Physics \&\ Astronomy\\
University of California\\ 
Riverside, CA 92521, USA}
\nocollaboration

\author{Fred Hamann}
\affiliation{Department of Physics \&\ Astronomy\\
University of California\\ 
Riverside, CA 92521, USA}
\nocollaboration

\author{Leah Simon}
\affiliation{Hands On Labs\\
750 W. Hampden Ave., Suite 100\\
Englewood, CO 80110, USA}
\nocollaboration

\author{Bo Ma}
\affiliation{School of Physics \&\ Astronomy\\
Sun Yat-Sen University\\ 
ZhuHai 519000, China}

\nocollaboration


\begin{abstract}

Rich complexes of associated absorption lines (AALs) in quasar spectra provide unique information about gaseous infall, outflows, and feedback processes in quasar environments. We study five quasars at redshifts 3.1 to 4.4 with AAL complexes containing from 7 to 18 \civ~\lam1548, 1551 systems in high-resolution spectra. These complexes span velocity ranges $\lesssim$3600~\kms\ within $\lesssim$8200~\kms\ of the quasar redshifts. All are highly ionised with no measurable low-ionisation ions like \siii\ or \cii, and all appear to form in the quasar/host galaxy environments based on evidence for line locking, partial covering of the background light source, strong \nv\ absorption, and/or roughly solar metallicities, and on the implausibility of such complexes forming in unrelated intervening galaxies. Most of the lines in all five complexes identify high-speed quasar-driven outflows at velocity shifts $v\lesssim -1000$~\kms. Four of the complexes also have lines at smaller blueshifted velocities that might form in ambient interstellar clouds, low-speed outflows or at feedback interfaces in the host galaxies where high-speed winds impact and shred interstellar clouds. The partial covering we measure in some of the high-speed outflow lines require small absorbing clouds with characteristic sizes $\lesssim$1~pc or $\lesssim$0.01~pc. The short survival times of these clouds require locations very close to the quasars, or cloud creation in situ at larger distances perhaps via feedback/cloud-shredding processes. The AAL complex in one quasar, J1008+3623, includes unusually narrow \civ\ systems at redshifted velocities $350\lesssim v\lesssim640$~\kms\ that are excellent candidates for gaseous infall towards the quasar, e.g., ``cold-mode'' accretion or a gravitationally-bound galactic fountain. 

\end{abstract}


\section{Introduction}

High-redshift quasars mark active stages of black hole growth and massive galaxy evolution at early cosmic times. Some galaxy evolution models suggest that quasars may follow a recent merger of gas-rich galaxies to trigger both rapid star formation and accretion onto the central supermassive black hole (SMBH), possibly with powerful outflows (feedback) driven by the central quasar and/or star formation \citep{Elvis06, Hopkins08, Veilleux09}. Quasar outflows play an important role in expelling gas and dust, which can quench star formation and regulate the growth of the central SMBH \citep{Kauffmann00, DiMatteo05, Hopkins10, Debuhr12}. Gaseous infall from the intergalactic medium (IGM) (e.g. cold mode accretion) is also important for fueling the central SMBH accretion and star formation during this active evolution phase \citep{Katz03, Keres09, Keres12, Tumlinson17}. It is possible that infall and outflow occur together if cold mode accretion is involved in triggering the star formation and quasar activity that drive the outflows \citep{Costa14, Nelson15, Suresh15}. However, in spite of these theoretical expectations, direct observational evidence for infalling gas is still lacking \citep{Prochaska09, Steidel10, Faucher11}. 

Associated absorption lines (AALs) in quasar spectra are valuable tools to study the gaseous environments of quasars and their host galaxies along direct lines of sight to the galactic centers. They can therefore uniquely measure the radial motions of gas into or out of the galaxies. AALs are defined by narrow velocity widths (less than a few hundred \kms), much different from broad absorption lines (BALs) that have velocity widths $\gtrsim$2000 \kms\ at speeds that can reach $>$0.1$c$ \citep{Anderson87, Weymann91, Hamann04, Simon10, Muzahid13}. While the broad widths and large velocity shifts of BALs clearly identify high-speed quasar-driven outflows, the origins of AALs are more diverse and often more ambiguous. The term `associated' means they have absorption redshifts are within several thousand \kms\ of the quasar emission-line redshift (i.e., $z_{abs}\approx z_{em}$, \citealt{Weymann79, Foltz86, Hamann97d}). AALs can have a range of physical origins that includes outflows very close to the quasars, interstellar gas in the host galaxies, infalling material from the intergalactic medium (IGM), and cosmologically intervening gas or galaxies unrelated to the quasars \citep{Sargent82, Tripp96, Hamann01, Odorico04, Hamann11}. 

We consider AALs to be `intrinsic' if they are physically related to the quasars, broadly within the influence of the quasar's radiation field \citep[e.g., at radial distances $\lesssim$1 Mpc for luminous quasars,][]{Perrotta18}. For luminous quasars, this correspond to radial distances hundreds of kpc up to $\sim$1 Mpc \citep{Wild08, Perrotta18}. Several observational tests have been proposed to determine if individual AAL systems are physically related to the quasars \citep[][]{Petitjean94, Barlow97b, Hamann97e, Hamann99, Ganguly99, Srianand00, Hamann01, Wise04, Narayanan04, Schaye07, Misawa07, Misawa07c, Arav05, Arav08, Hamann11, Simon12}. They include 1) absorption line variability, 2) partial covering of the quasar light source that requires very small absorbing clouds, 3) high gas densities inferred from excited-state absorption lines, and 4) line profiles indicative of gas flows because they are broad and smooth compared to thermal velocities and nominal intervening (unrelated) absorption lines \citep{Rauch98}. Supersolar metallicities and `line-locking' between pairs of doublet lines are also consistent with formation in outflows from a metal-rich galactic nuclear regions \citep{Ganguly03, Prochaska06, Schaye07, Simon10, Hamann11, Bowler14}. Line locks, in particular, are believed to identify situations where radiation pressure drives the outflow \textit{and} the locked lines play a significant role in the photon momentum transfer. 

Previous studies indicate that AALs have a high probability of being intrinsic. For example, statistical tests by \citet{Richards99} and \citet{Richards01a} argued that around 36\% of high-velocity \civ\ absorbers tend to be intrinsic based on a correlation between the numbers of these absorbers detected and the radio properties of the quasars. \citet{Misawa07} used a covering fraction analysis to show that $\gtrsim$33\% of quasars have intrinsic \civ\ systems within 5000 \kms\ of the quasar emission redshift \citep[see also][]{Simon12, Perrotta16, Perrotta18}. Much larger statistical studies by \citet{Nestor08} and \citet{Wild08} indicate that $\gtrsim$43\% of \civ\ AALs with rest equivalent width REW$\geq$0.3 \AA\ in the velocity range from $\sim$750 to $\sim$12,000 \kms\ form in quasar-driven outflows. This minimum outflow fraction peaks at $\sim$80\% for AAL velocity shifts near 2000 \kms. 

In this study, we examine rich multi-component complexes of AALs that provide unique constraints on the nature of gas flows in quasar environments during the active quasar stages of massive galaxy evolution. For example, one interesting possibility is that AAL complexes are direct signatures of feedback where fast quasar-driven outflows impact, shred, and disperse interstellar clouds in the host galaxies to produce many discrete absorption-line clumps \citep{Hopkins10, Faucher11}. Rich AAL complexes might also form in unusual highly-structured quasar outflows without (or before) interactions with gas in the host galaxies \citep{Morris86, Dobrzycki99, Ganguly03, Misawa03, Misawa05, Misawa07, Simon10b, Hamann11}. For AALs at redshifts $z_{abs} > z_{em}$, a third possibility is that they identify gas condensations in cold-mode accretion from the intergalactic medium. In particular, \citet{McCourt16} argue that infalling clouds that are optically-thin and pressure-confined will naturally fragment as they cool, finally reaching very small sizes $\sim$0.1 pc across. 

Our study extends the recent work by \citet{Simon10b} and \citet[][hereafter C18]{Chen18} that also use high-resolution spectra to analyze AALs and AAL complexes in bright quasars at redshifts $\gtrsim$2. Here we examine five quasars selected specifically to have rich AAL complexes based on the \civ\ \lam 1548, 1551 absorption lines. We describe the quasar sample and spectra in Section 2. We identify and fit the lines to measure the gas kinematics, column densities and covering fractions in Section 3. We analyze additional physical properties such as the ionisation and metallicity in Section 4, and we discuss the nature and origins of these complexes in Section 5. We conclude with a brief summary in Section 6. Throughout this paper, we adopt a cosmology with $H_0=71$ \kms\ Mpc $^{-1}$, $\Omega_M=0.27$ and $\Omega_{\Lambda}=0.73$.

\section{Data Overview}

\subsection{Quasar Sample}

We select five quasars for our study from the larger sample of 24 AAL quasars in \citet{Simon12} and \citet{simonphd}. They obtained high-resolution spectra at the W. M. Keck Observatory (Keck), the Magellan Clay Telescope (Magellan), or the ESO Very Large Telescope (VLT). We choose these particular quasars based on the presence of rich \civ\ AAL complexes, which we define throughout this paper as groups of more than $\sim$6 \civ\ AALs spread over $\lesssim3000$ \kms\ and within $\sim$8000 \kms\ of the emission-line redshift that appear related to each other based on similar velocity shifts, similar profiles, partial covering of the background light or other evidence of an outflow/intrinsic origin \citep[see also][]{simonphd}. The quasars in our sample, their redshifts, and some basic information about the spectra are listed in \Cref{tab:sample}. 

\begin{figure}
\centering
\includegraphics[width=0.7\textwidth]{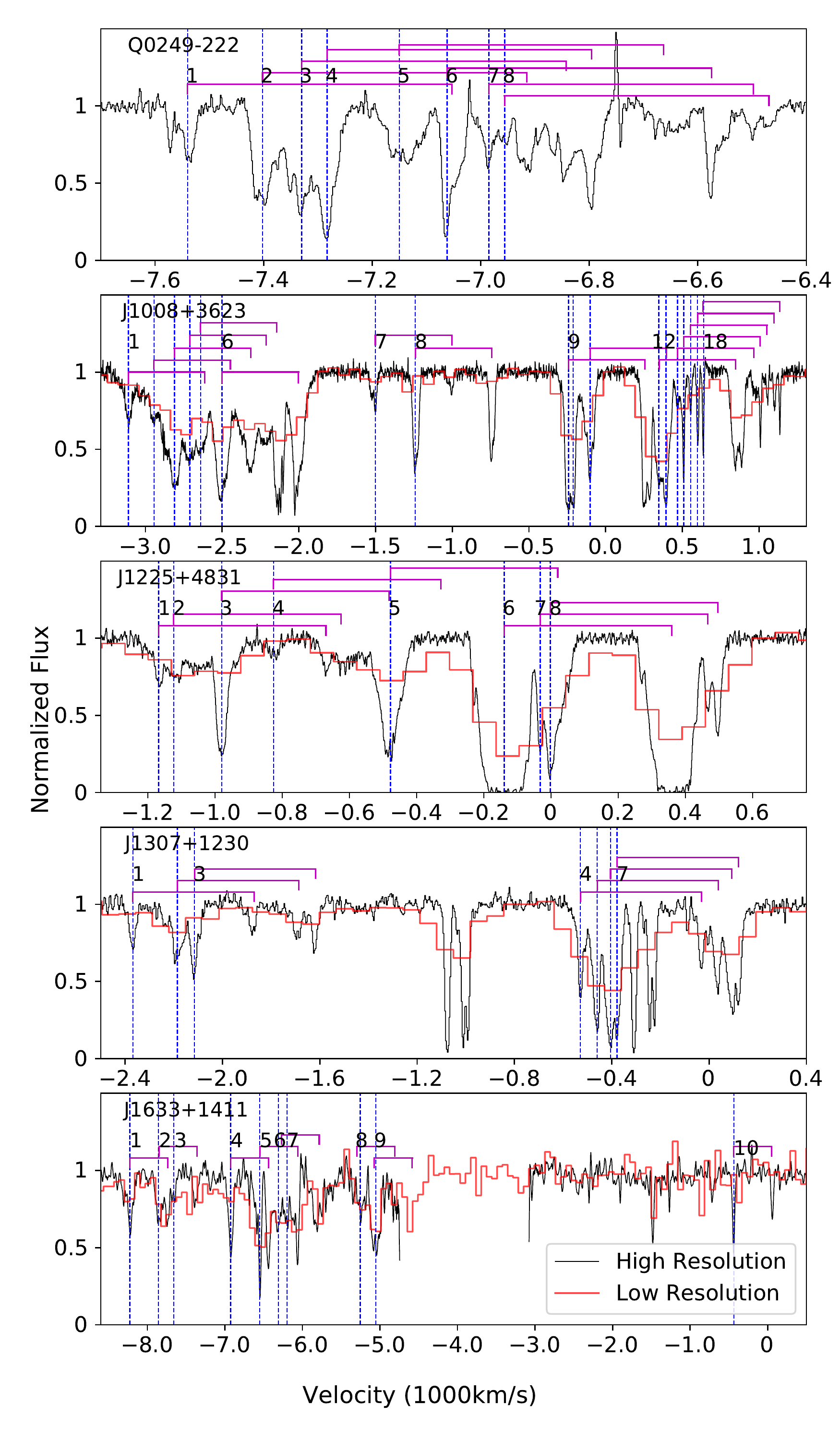}
\caption{Normalized \civ\ line profiles of the five quasars in our sample plotted on a velocity scale relative to the quasar redshift (\Cref{tab:sample}). The high-resolution spectra by our team are shown in black, and the low-resolution spectra from SDSS are shown in red (grey solid). The blue dash lines are identified AALs based on the high-resolution spectra, and the brackets show the doublets. The velocities pertain to the short-wavelength lines in the doublets. The quasar Q0249$-$222 does not have a spectrum from SDSS.\label{fig:SDSS_com}}
\end{figure}

\Cref{fig:SDSS_com} shows the \civ\ AAL complexes in the high-resolution we analyze compared to low-resolution spectra from the Sloan Digital Sky Survey \citep[SDSS,][]{York00} when available. This comparison shows that high-resolution spectra are essential to identify and study the AAL complexes. Other quasars with AAL complexes in the \cite{Simon12} and \citet{simonphd} sample (Q0334$-$204, Q2044$-$168, and J102325.31+514251.0) are discussed in C18 or \citet{Simon10b}.

\subsection{Observations and Data Reductions}

\Cref{tab:sample} summarizes basic information about the data and observations. The spectra of J100841.22+362319.3, J122518.66+483116.3 and J163319.62+141142.0 (hereafter J1008+3623, J1225+4831 and J1633+1411) were obtained with High Resolution Echelle Spectrometer (HIRES) on Keck, the spectra of Q0249$-$222 were observed with UV-Visual Echelle Spectrograph (UVES) on the VLT, and the spectra of J130710.25+123021.6 (hereafter J1307+1230) were observed with Magellan Inamori Kyocera Echelle (MIKE) spectrograph on Magellan. All of these spectra cover important absorption lines from at least \lya\ \lam 1216 to \civ\ \lam 1548, 1551 in the quasar rest frame. We used standard techniques to reduce the spectra. In particular, for the Keck-HIRES spectra, we used a software package MAKEE \citep[as described in][]{Barlow97, Hamann97} for initial data reduction and spectral extraction. These procedures are described in detail in \citet{Hamann01}. The VLT-UVES data reduction procedures are described in \citet{simonphd}. We reduce the Magellan-MIKE data using the Mike-Redux IDL\footnote{Interactive Data Language (IDL) is a programming language used for data analysis.} routines package \citep[as described in][]{Bernstein15}. All of the spectra are shifted to vacuum heliocentric wavelengths. 

We use the IRAF\footnote{Image Reduction and Analysis Facility (IRAF) is maintained and distributed by the National Optical Astronomy Observatories.} software package for additional data processing. In particular, we normalize the reduced spectra to unity by fitting a pseudo-continuum to each quasar spectrum, including the emission lines, constrained by regions that appear free of absorption. The fitting function is a low-order polynomial. For spectral regions where the continuum is affected by significant absorptions, e.g., in the \lya\ forest, we visually inspect the spectra to find small segments of continuum not affected by absorption or obvious noise spikes and, again, fit with a low-order polynomial.

We also obtained near-IR spectra of the quasars J1008+3623 and J1307+1230 with Gemini Near Infra-Red Spectrograph (GNIRS) in the cross-dispersed mode with the short camera, the 111 l/mm grating and a 0.3 arcsec slit providing resolution of $\sim$51 \kms\ on Gemini North Telescope in the K band covering the \hb-[\oiii] region in January 2016. The goal of these observations was to measure the quasar redshifts from their [\oiii] \lam 4959, 5007 emission lines. We followed the standard procedures to reduce these spectra \citep[as described in][]{Cooke05} using the Gemini IRAF package \footnote{http://www.gemini.edu/sciops/data-and-results/processing-software}.

\subsection{Redshift Estimates}

Accurate quasar-frame redshifts are important to measure flow speeds and test for infall versus outflow in the AAL complexes. Redshifts derived from the UV broad emission lines are known to be uncertain because of gas motions/outflows in the broad emission-line regions. High-ionisation lines like \civ\ \lam 1548, 1551 are typically blueshifted by several hundred \kms\ compared to the redshift determined from narrow forbidden lines such as [\oiii] \lam 5007, which is generally regarded to be the best UV/optical indicator of quasar systemic redshifts \citep{Gaskell82, Shen07, Wang11}. Low-ionisation permitted lines such as \mgii\ \lam 2800 can be good redshift indicators because their velocity shifts relative to [\oiii] are typically $\lesssim200$ \kms\ \citep{Richards02, Shen07, Shen16b}. 

Unfortunately, our Gemini-GNIRS observations (Section 2.2) designed to measure redshifts from the [\oiii] emission lines in J1008+3623 and J1307+1230 yielded only non-detections but at high significance. The absence of these lines might have important implications for the quasar environments and the nature of the AAL complexes (Section 5.3). However, without [\oiii] measurements, we search the literature to find the best available redshift for every quasar in our study. The results are listed in \Cref{tab:sample}. For J1008+3623, J1225+4831, J1633+1411 and J1307+1230, we adopt redshifts from \citet{Hewett10} based on UV emission lines in the SDSS spectra. These redshifts are typically displaced blueward by $\sim$100 \kms\ from [\oiii]-based redshifts \citep{Shen16, Shen16b}. We adopt redshifts from \citet{Hewett10} and recalculate their uncertainties to include this typical offest, $\sim$100 \kms. We also visually inspect the SDSS spectra to check for redshift anomalies based on the \cii\ \lam 1335, \feii\ \lam 1786, or \ciii] \lam 1909 emission lines. We find none; the redshifts listed in \Cref{tab:sample} provide a good match to the low-ionisation emission lines present in the spectra. For the quasar Q0249$-$222, which is not in the SDSS database, we adopt the emission-line redshift and uncertainty from \citet{Tytler92} who apply a maximum-likelihood measurement technique to the emission lines of \lya\ \lam 1216, \civ\ \lam 1548, 1551 and \siiv+\oiv] \lam 1398.

\section{Absorption-line Measurements}

\Cref{fig:Q0249,fig:J1008,fig:J1225,fig:J1307,fig:J1633} provide a complete view of all detected AALs belonging to the complexes in each quasar in our study. The spectra are plotted on a velocity scale relative to the emission-line redshifts listed in \Cref{tab:sample}. The spectra are also normalized to unity in the continuum using smooth curves constrained to match the observed flux in narrow wavelength regions free of absorption lines.  

We identify individual velocity components in the AAL complexes starting with the \civ\ \lam 1548, 1551 doublet because it is not contaminated in \lya\ forest and \civ\ tends to be strong at wavelengths with high signal-to-noise ratios in our spectra. We then search for a wide range of other plausible lines at these same redshifts, such as \svi\ \lam 933, 944, \cii\ \lam 1335, \ciii\ \lam 977, \siii\ \lam 1260, 1304, 1527, \siiii\ \lam 1207, \siiv\ \lam 1394, 1403, \niii\ \lam 990, \nv\ \lam 1239, 1243, \ovi\ \lam 1032, 1038 and Lyman lines. We also search for various resonance/excited-state line pairs, such as \siii/\siii*, \cii/\cii*, \ciii/\ciii*, \niii/\niii* and \siv/\siv*.  We have also reviewed our assessments of upper limits on the excited-state lines and find that none of them are useful for density and location constraints (see C18 for more discussion). The lines identified in each quasar are listed in \Cref{tab:Q0249,tab:J1008,tab:J1225,tab:J1307,tab:J1633} below. 

\begin{figure}
\centering
\includegraphics[width=0.95\textwidth]{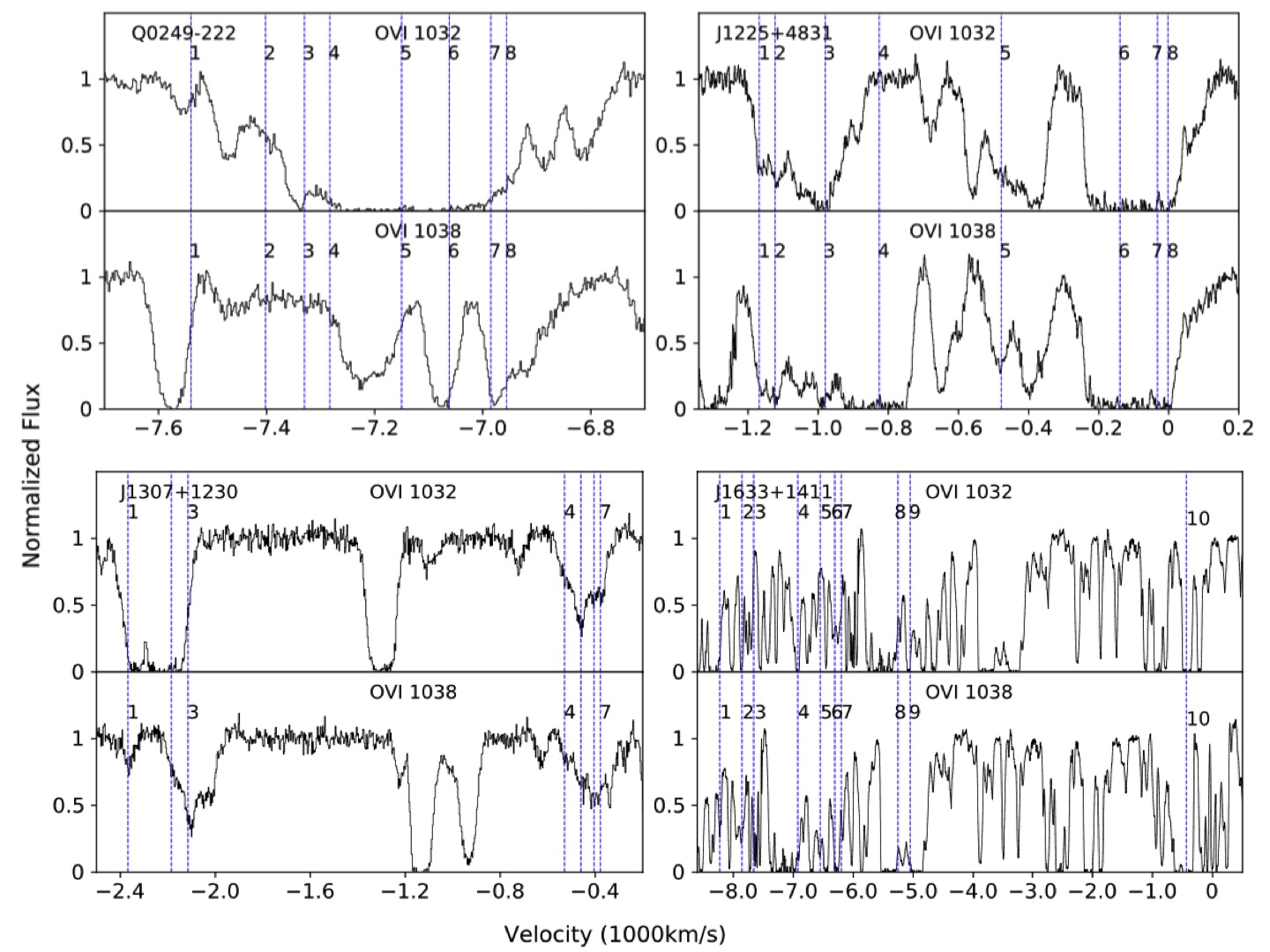}
\caption{Normalized \ovi\ \lam 1032, 1038 line profiles of Q0249$-$222, J1225+4831, J1307+1230 and J1633+1411 in our sample plotted on a velocity scale relative to the quasar redshift (\Cref{tab:sample}). The blue dash lines mark the locations of identified AALs based on the \civ\ lines. All these \ovi\ lines are not clearly present.\label{fig:ovi}}
\end{figure}

It is necessary to note that we also list the lines that are clearly present but not measurable due to blending/contamination (e.g., in the \lya\ forest) in the tables without measurements, and we do not provide their upper limits because they are too large to be useful for further analysis. For \ovi\ lines, particularly, they might be present in all the five quasars, but we cannot make reliable assessments for Q0249$-$222, J1225+4831, J1307+1230 and J1633+1411 because of blending/contamination problems in the \lya\ forest. Since \ovi\ is an important outflow line, we have therefore created a separate plot \Cref{fig:ovi} showing these *possible* \ovi\ detections that are not clearly present in these four quasars. We now describe the line measurements and the contents of the data tables.

\subsection{Line Fit Procedures}

We follow the line-fitting procedures described in C18. This involves fitting each AAL component with a Gaussian optical depth profile convolved with a Gaussian kernel that represents the instrumental broadening. Specially, the Gaussian optical depth profile is defined by
\begin{equation}
\label{eq:1}
\tau_{v}=\tau_{0}e^{-v^2/b^2},
\end{equation}
where $\tau_v$ is the optical depth at velocity $\textrm{v}$ and $b$ is the Doppler parameter. $\tau_{0}$ is the line-center optical depth related to the lower-state/ion column density, $N$, by 
\begin{equation}
\label{eq:2}
\tau_{0}=\frac{\sqrt{\pi}e^2}{m_{e}c}\frac{Nf\lambda_{0}}{b},
\end{equation}
where $f$ is the oscillator strength, and $\lambda_{0}$ is the laboratory wavelength. We assume for simplicity that the background light source has a uniform brightness, the absorbing medium is homogeneous (with the same optical depth along every line of sight), so the observed intensity at velocity $\textrm{v}$ is 
\begin{equation}
\label{eq:3}
\frac{I_{v}}{I_{0}}=1-C_v+C_ve^{-\tau_{v}},
\end{equation}
where $I_{0}$ is the continuum intensity, $I_{v}$ is the measured intensity, and $C_v$ is the covering fraction of the absorbing medium across the emission source with values $0 < C_v \leq1$ \citep{Ganguly99, Hamann97b, Barlow97b}. We also assume that the covering fraction is constant with velocity across the line profiles such that $C_v \equiv C_0$, which is a good approximation for the narrow lines in our study. 

We use four different fitting procedures depending on the circumstances of each line (see C18 for more discussion). Briefly, these procedures involve 1) Heavily saturated lines based on $\sim$1:1 doublet ratios or flat-bottom profiles (see for example components 2, 3, 4, 5, 6 and 7 of \lya\ in \Cref{fig:Q0249} and component 6 of \civ\ \lam 1548, 1551 in \Cref{fig:J1225}). In these situations, the covering fraction equals the normalized depth of the line, and we derive lower limits on the column densities based on a conservative minimum optical depth of $\tau_0\gtrsim3$ in the weakest doublet/multiplet component. 2) Doublets or multiplets that appear unsaturated based on intermediate line ratios, e.g., between 2:1 and 1:1 for the doublets\footnote{The optical depth ratios for lines sharing a common lower energy state are set by the ratio of their $f\lambda$ values, where $f$ is the oscillator strength and $\lambda$ is the line wavelength. This ratio is $\sim$2:1 for the doublets discussed in this paper, such as the optical depth ratio of \civ\ \lam 1548 and \lam 1551.} (such as components 3 and 4 of \civ\ \lam 1548, 1551 in \Cref{fig:J1008}). In this case, we solve for both $C_0$ and $\tau_0$ by fitting the doublet lines simultaneously. This procedure can also apply to multiple lines in the Lyman series. 3) Weak doublets that appear to have $\tau_0\ll 1$ based on $\sim$2:1 strength ratios (such as components 14 and 16 to 18 of \civ\ \lam 1548, 1551 in \Cref{fig:J1008}). In this case, the values of $\tau_0$ and $C_0$ in \Cref{eq:1,eq:3} are degenerate and cannot be determined separately. Thus we adopt $C_0=1$ to derive $\tau_0$ and the column density values that are lower limits if the actual absorber has $C_0<1$. And 4) single lines that do not reach zero intensity (such as component 9 of \civ\ \lam 1548 in \Cref{fig:J1633} and components 1 to 6 of \nv\ \lam 1243 in \Cref{fig:J1008} where the other doublet components \civ\ \lam 1551 or \nv\ \lam 1239 are not measured). Here again we adopt $C_0=1$ to derive lower limits on $\tau_0$ and the column densities. 

\subsection{Line Fit Results}

\begin{figure}
\centering
\includegraphics[width=0.6\textwidth]{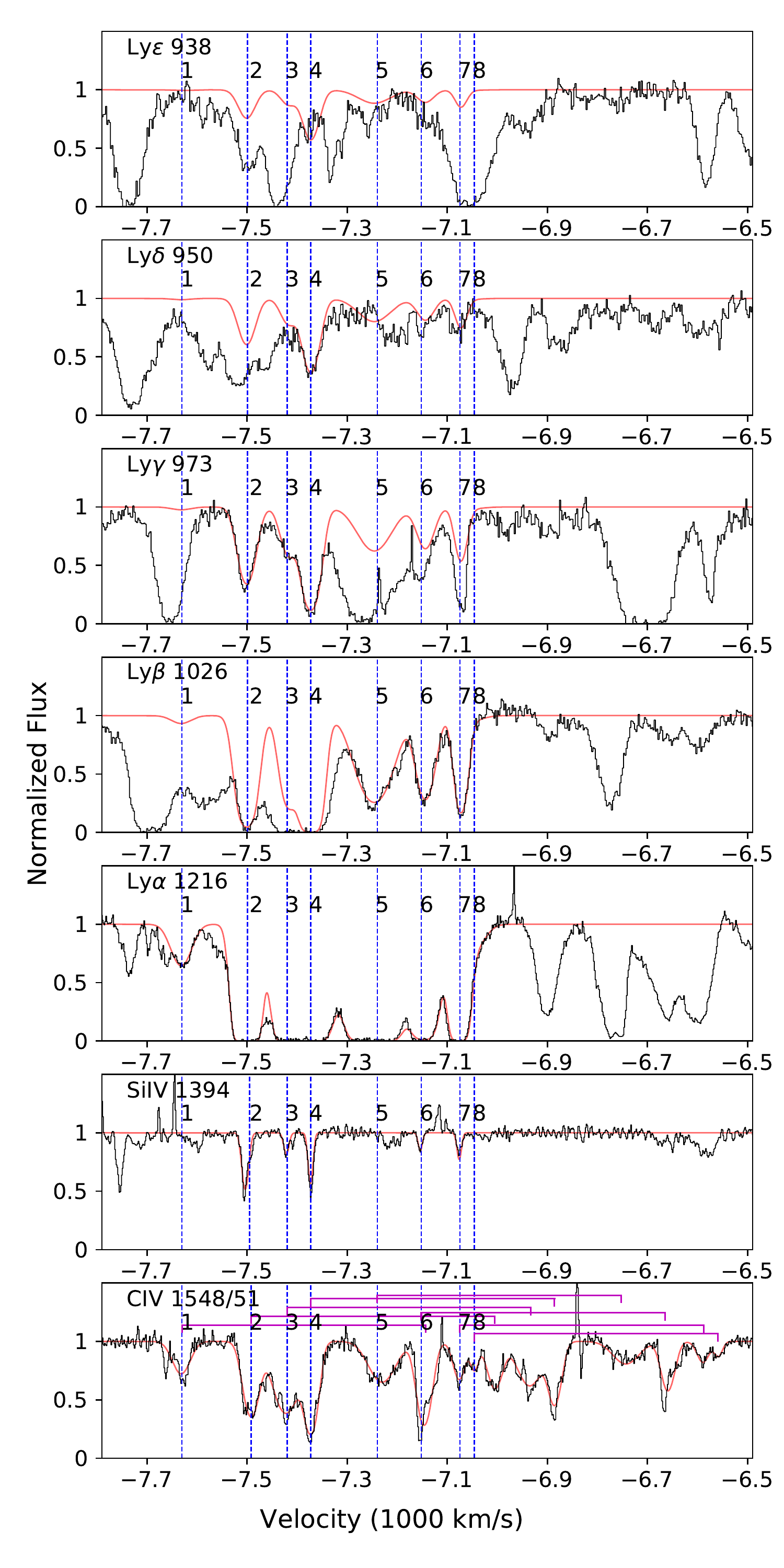}
\caption{Q0249$-$222: Normalized line profiles in the VLT-UVES spectra plotted on a velocity scale relative to the quasar redshift (\Cref{tab:sample}). The spectra are shown in black, and the final fitting lines are shown in red (grey solid). The blue dash lines are identified components from 1 to 8, and the brackets show the doublets. The velocities pertain to the short-wavelength lines in the doublets.\label{fig:Q0249}}
\end{figure}

\begin{figure*}
\centering
\includegraphics[width=1.0\textwidth]{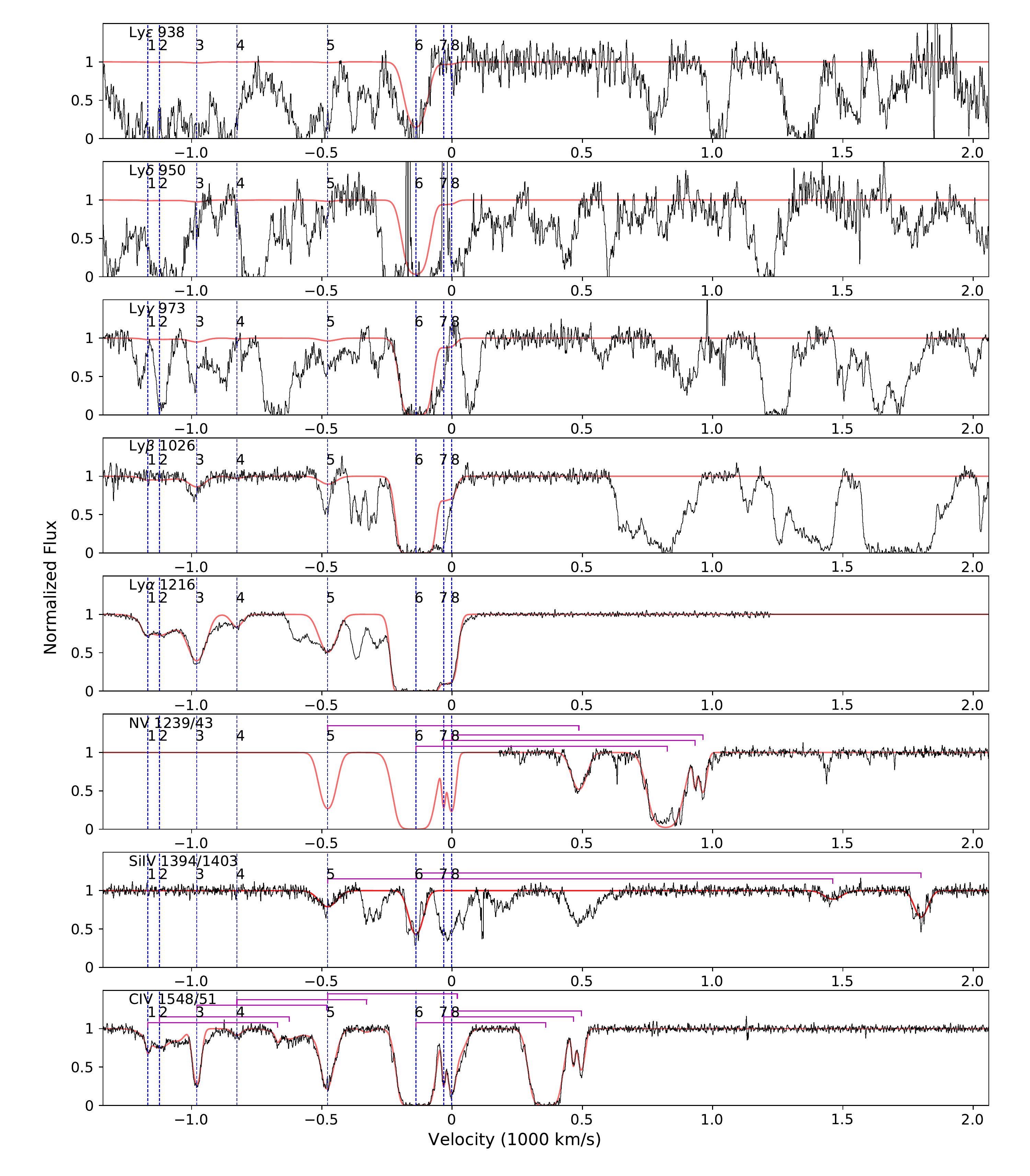}
\caption{J1225+4831: Normalized line profiles in the KeckI-HIRES spectra plotted on a velocity scale relative to the quasar redshift (\Cref{tab:sample}). The spectra are shown in black, and the final fitting lines are shown in red (grey solid). The blue dash lines are identified components from 1 to 8, and the brackets show the doublets. The velocities pertain to the short-wavelength lines in the doublets.\label{fig:J1225}}
\end{figure*}

\begin{figure*}
\centering
\includegraphics[width=1.0\textwidth]{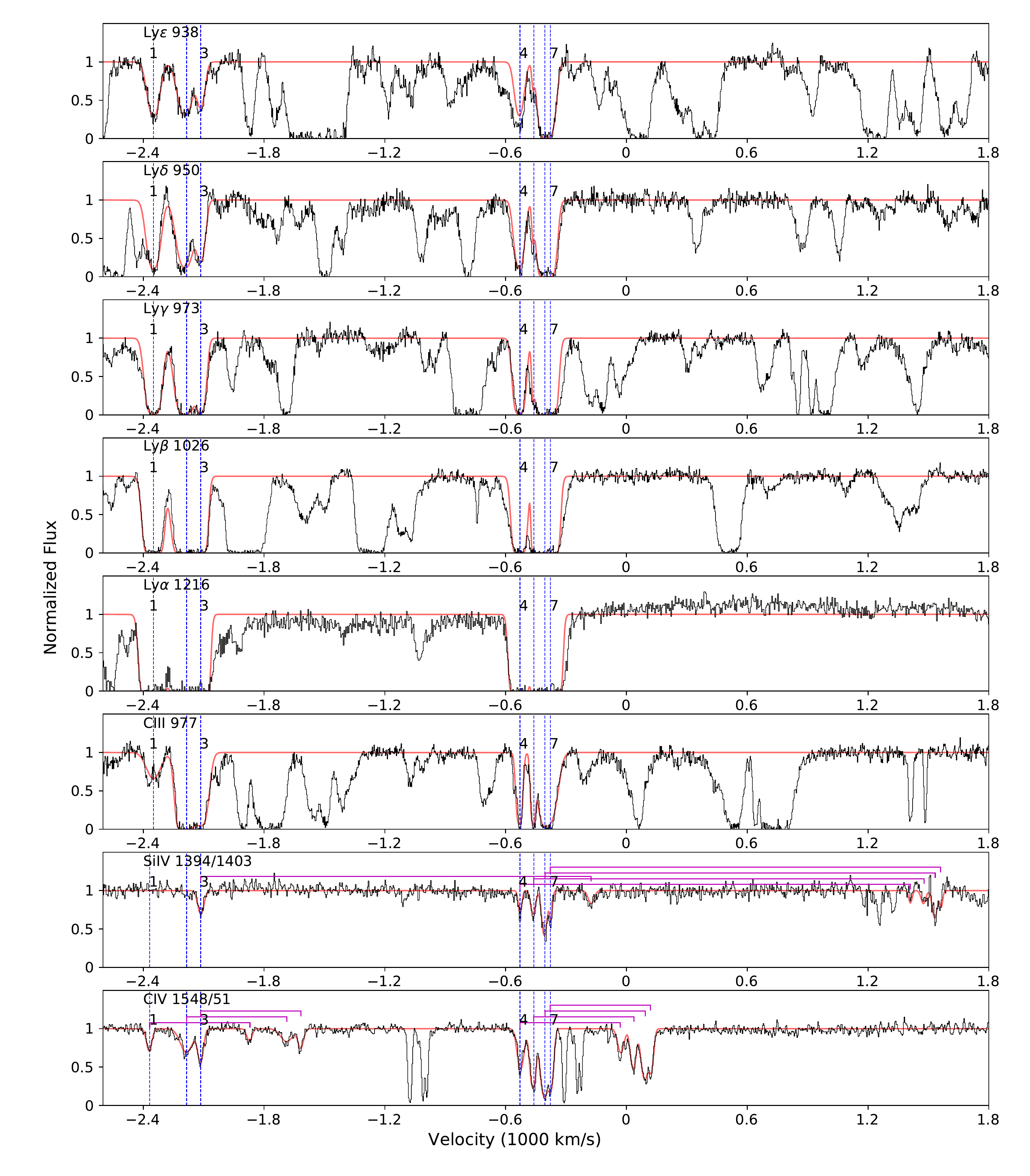}
\caption{J1307+1230: Normalized line profiles in the Magellan-MIKE spectra plotted on a velocity scale relative to the quasar redshift (\Cref{tab:sample}). The spectra are shown in black, and the final fitting lines are shown in red (grey solid). The blue dash lines are identified components from 1 to 7, and the brackets show the doublets. The velocities pertain to the short-wavelength lines in the doublets.\label{fig:J1307}}
\end{figure*}

\begin{figure*}
\centering
\includegraphics[width=0.95\textwidth]{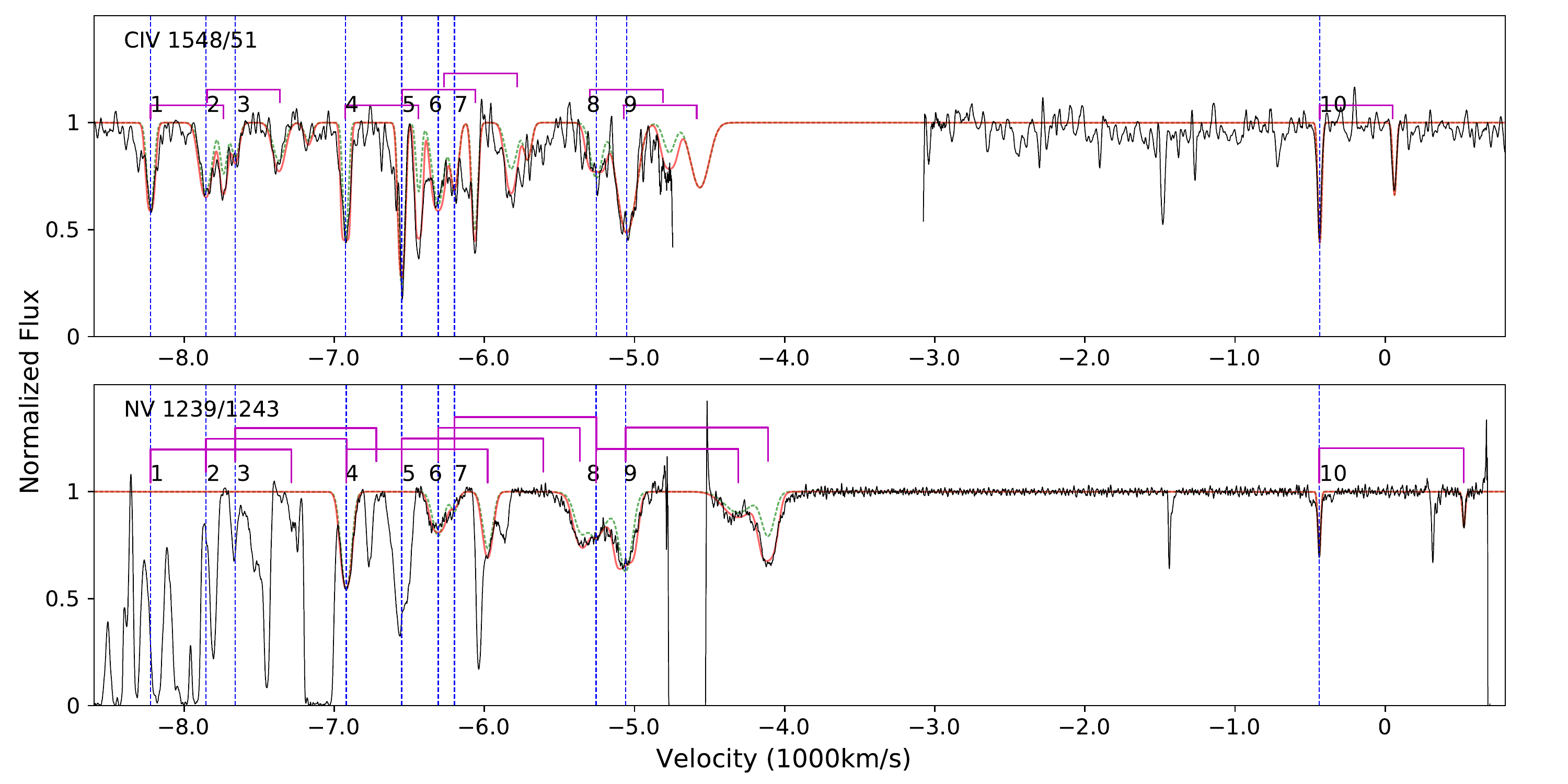}
\caption{J1633+1411: Normalized line profiles in the KeckI-HIRES spectra plotted on a velocity scale relative to the quasar redshift (\Cref{tab:sample}). The spectra are shown in black, and the final fitting lines are shown in red (grey solid). The blue dash lines are identified components from 1 to 10, and the brackets show the doublets. The velocities pertain to the short-wavelength lines in the doublets. The green dash lines show the predicted lines based on \civ\ \lam 1548, or \nv\ \lam 1239 assuming $C_0=1$.\label{fig:J1633}}
\end{figure*}

The final line fits are shown by the red (grey solid) curves in \Cref{fig:Q0249,fig:J1008,fig:J1225,fig:J1307,fig:J1633}. \Cref{tab:Q0249,tab:J1008,tab:J1225,tab:J1307,tab:J1633} list the fit parameters and associated uncertainties. Notes in the last column provide information on blends or additional uncertainties. Footnotes in the tables indicate the fitting procedure used. The line data are organized in the tables according to the redshifts, i.e., component numbers in the first column as labeled in \Cref{fig:Q0249,fig:J1008,fig:J1225,fig:J1307,fig:J1633}. The uncertainties listed for most of the parameters are 1$\sigma$ errors derived from the fits, affected mainly by pixel-to-pixel noise fluctuations in the spectra. They do not consider errors in the continuum placement. For very optically thick or/and blended lines, we follow the same procedures described in C18 to obtain direct estimates or limits on the parameter values. For the components that show multiple Lyman lines, we try to fit as many Lyman lines as possible simultaneously to get the best constraints on the column densities $N$(\hi) and covering fractions $C_0$.

We find partial covering in many AAL components in the quasars J1008+3623 and J1633+1411. In particular, components 9 and 10 in \civ\ \lam 1548, 1551 and \nv\ \lam 1239, 1243, component 6 in \ovi\ \lam 1032, 1038 of J1008+3623, and component 9 in \nv\ of J1633+1411 appear to have $\sim$1:1 doublet ratios at strengths that do not reach zero intensity. Other lines reveal partial covering on closer examination. \Cref{fig:J1008,fig:J1633} show how fits to the stronger short-wavelength lines assuming $C_0=1$ in some doublets will under-predict the strength of the weaker long-wavelength doublet lines. In particular, for J1008+3623 in \Cref{fig:J1008} see components 6, 9 and 10 in the \ovi\ doublets, components 9, 10 and 11 in \nv, and components 3, 4, 9 and 10 in \civ, and for J1633+1411 in \Cref{fig:J1633} see components 1 to 6 in \civ, and components 4 and 9 in \nv . The predicted long-wavelength doublet lines assuming $C_0=1$ (shown by the green dash curves) are too weak in all cases, indicating that these features have $C_0 <1$ with specific derived values listed in \Cref{tab:J1008,tab:J1633}. We cannot determine if there is partial covering in other lines of J1008+3623 and J1633+1411 because they are either blended, weak, or only one line in the doublet pair is within our wavelength coverage (see notes in \Cref{tab:J1008,tab:J1633}). There is no partial covering in any of the AALs measured in the other quasars, Q0249$-$222, J1225+4831, and J1307+1230.

\section{Analysis}

\subsection{Photoionisation Models}

We assume all of the absorbers in our sample are photoionised by the quasar spectrum, close enough to the quasars to be within their radiative sphere of influence \citep[][see Section 1 for more information]{Wild08, Perrotta18}. We perform a photoionisation analysis of selected components in the AAL complexes that have well-measured lines in multiple ions that can constrain the ionisation, such as \siiv, \civ, \nv, \ovi, and the \hi\ Lyman series (see more discussion in C18). We specifically examine components 3, 4, 6, and 7 in Q0249$-$222, components 1 to 4, 6, 8 to 13, and 15 to 18 in J1008+3623, components 5 to 8 in J1225+4831, components 1, and 4 to 7 in J1307+1230, and components 4 and 10 in J1633+1411. We use the photoionisation code \cloudy\ \citep{Ferland17} to estimate the ionisation, total column density, and element abundances. We run \cloudy\ using a generic fixed \hi\ column density of $\log N(\hi)$(\cmN)~$=15$ that ensures that the clouds are optically thin in the Lyman continuum. This is justified by the measurements and upper limits on $N(\hi )$ that we derive from the data (\Cref{tab:Q0249,tab:J1008,tab:J1225,tab:J1307}). In this optically thin regime, the column density ratios needed for our analysis do not depend on $N(\hi)$ nor $N$(H). The calculations assume solar abundances and a standard quasar ionising spectrum. The quasar spectrum and other details of the calculations are described in C18. \Cref{tab:parameter} lists the main results, which we describe below.  

\subsubsection{Ionisation \& Total Column Densities}

\begin{figure*}
\centering
\includegraphics[width=0.8\textwidth]{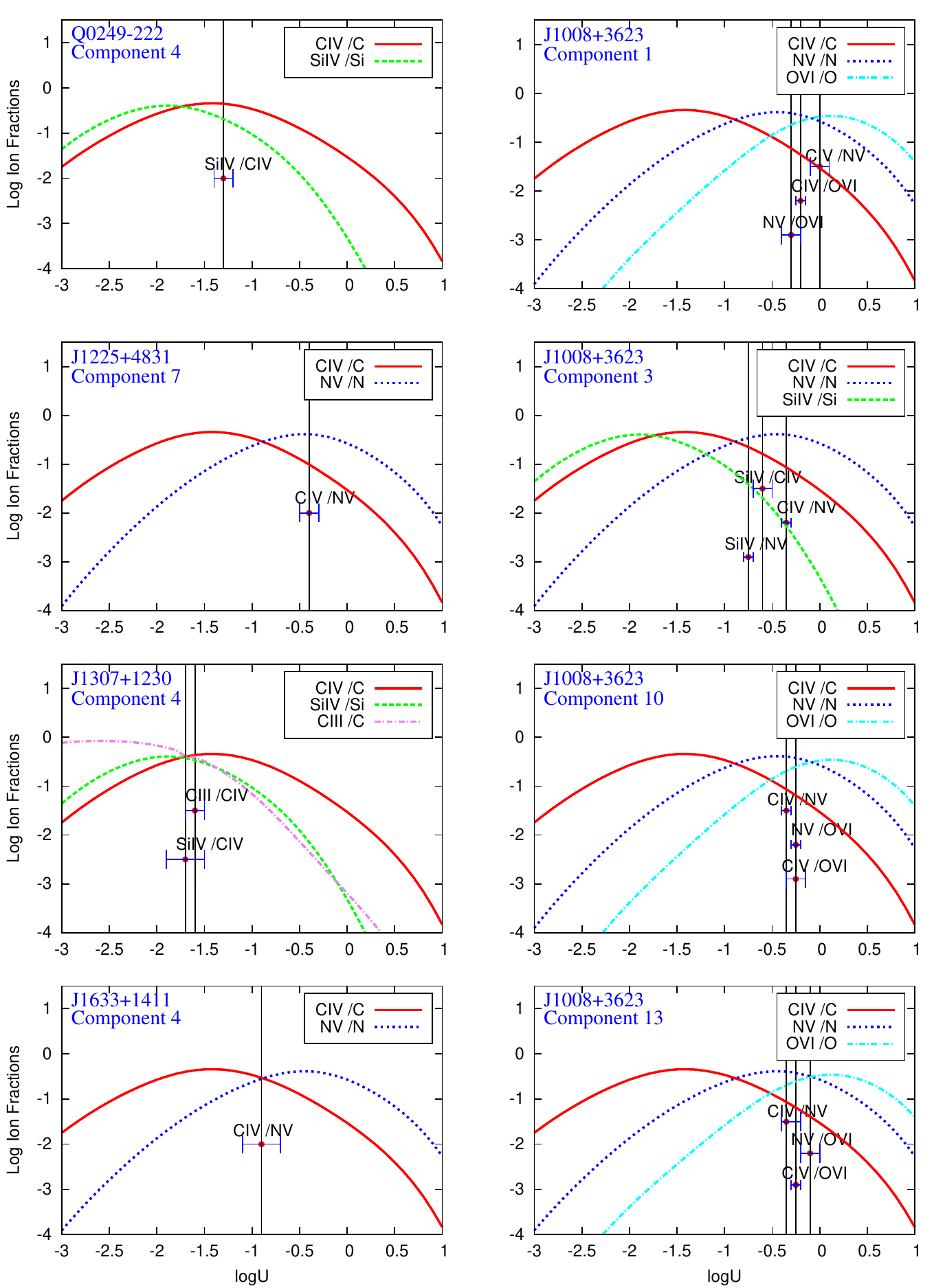}
\caption{Theoretical ionisation fractions, $f(M_i)$, for selected stages of the elements Si, C, N and O plotted against ionisation parameters $\log U$. The black vertical lines with blue error bars are the best estimations of $U$ for each ion pair. Eight panels show example components in five quasars.\label{fig:cloudy}}
\end{figure*}

The ionisation state of the absorbers is described by the ionisation parameter, $U$, which is the dimensionless ratio of the number density of hydrogen-ionising photons at the illuminated face of the clouds to the number density of hydrogen atoms. \Cref{fig:cloudy} shows examples of the calculated ionisation fractions in important ions compared to well-measured column density ratios in the data (e.g., \siiv/\civ, \civ/\nv, \nv/\ovi, etc.). These comparisons yield estimates of the ionisation parameter $U$ (shown by vertical lines in the figures) with uncertainties (horizontal bars) based on the column density uncertainties listed in the data tables. 

Previous studies have shown that individual AAL systems can span a wide range of ionisations from \mgi, \siii, or \cii\ up to \nv, \ovi\ and even \neviii\ that require a wide range of $U$ values in the absorbing gas \citep[e.g.,][C18]{Hamann95, Petitjean99, Hamann01, Dunn10, Arav13, Muzahid13, Finn14}. None of the absorbers in our samples exhibits low-ionisation species such as \siii, \cii, or \mgii. Only higher ions like \siiv, \civ, or \nv\ are detected. We use column density ratios in these ions to obtain a single best $U$ value for each system. These best $U$ values, listed in \Cref{tab:parameter}, derive from a simple average of results from different column density ratios weighted by the measurement errors in those ratios (e.g., the horizontal bars in \Cref{fig:cloudy}). 

It is necessary to note that there is another uncertainty in $U$ due to uncertainties in the continuum shape we adopted when we ran \cloudy. Generally, we found the ionisation parameter $\log U$ in the models changed by 0.1 if considering 1$\sigma$ deviation in temperature, which is the most important parameter determining the continuum shape in the far-UV (see C18 for more details). We take this uncertainty into account, and list the final errors in \Cref{tab:parameter}, including both the measurement errors and the theoretical uncertainties in our \cloudy\ analysis.

We describe the detailed procedures for error estimates of $\log U$. \Cref{fig:cloudy} shows error bars from the measurement errors in various ion pairs based on an adopted spectral energy distribution (SED) with a temperature 350,000 K (see C18 for more details). For each system, we derive a weighted mean to be the single best $U$ value and a weighted error to be the measurement error. Then we ran additional \cloudy\ models with the temperatures 200,000 K and 500,000 K, which could be considered as 1 $\sigma$ deviations from the continuum we adopted. We measure the offsets of the ionisation parameter $\log U$ in these temperatures, and revise the measurement errors. The combined errors of $\log U$ are listed in \Cref{tab:parameter}. It is also important to note that we adopt a single $U$ for each system. But for a few systems such as component 3 in J1008+3623 (\Cref{fig:cloudy}), the ionization parameter $\log U$ extracted from the \siiv/\nv\ ratio is more than 3 $\sigma$ different than the one derived from the \civ/\nv\ ratio. This is potentially an interesting result, because it is common for AAL systems to exhibit a range of ionization parameters as explained above. However, we do not trust this difference in $\log U$ is real. We attribute this problem to line fits (see \Cref{fig:J1008}). Note that the \nv\ \lam 1239 line is not fitted but simply drawn because it falls in a gap in the wavelength coverage. The error bars shown in \Cref{fig:cloudy} do not capture these difficulties; they reflect only the formal errors returned by the fitting code that is fitting numerous lines simultaneously. They could be larger than indicated in \Cref{fig:cloudy} and therefore, we do not attach any significance to the different $\log U$ values indicated in \Cref{fig:cloudy}. 

We estimate the total hydrogen column density, $N(\rm{H})$, for each absorber by applying an ionisation correction, \hi /H, appropriate for the derived $U$ values to the $N$(\hi ) values measured from the data (Section 3). For some absorbers with only upper limits on $N$(\hi ), we derive upper limits on $N$(H). We could not constrain $N(\rm{H})$ for the absorbers in J1633+1411 because its Lyman lines are severely blended in the forest and could not be distinguished from unrelated lines.

\subsubsection{Gas Metallicity}

We estimate the absorber metallicities using the general relation
\begin{equation}
\left[\frac{\mbox{M}}{\mbox{H}}\right]=\log\left(\frac{N(\mbox{M}_i)}{N(\hi)}\right)+\log\left(\frac{f(\hi)}{f(\mbox{M}_i)}\right)+\log\left(\frac{\mbox{H}}{\mbox{M}}\right)_{\odot},
\end{equation}
where $(\mbox{H/M})_{\odot}$ is the solar abundance ratio of hydrogen to some metal $\mbox{M}$, $N(\hi)$ and $f(\hi)$ are the column density and ionisation fraction in \hi, respectively, and $N(\mbox{M}_i)$ and $f(\mbox{M}_i)$ are the column density and ionisation fraction in ion $\mbox{M}_i$ of metal $\rm{M}$, respectively. We use our estimates of $U$ from Section 4.1.1 to determine the ionisation correction $f(\hi )$/$f(\rm{M}_i)$ from the \cloudy\ calculations (as in \Cref{fig:cloudy}), and then plug in measured values of the column densities to obtain the metal abundance [M/H]. The results for [Si/H], [C/H], [N/H] and/or [O/H], depending on lines available, are listed in \Cref{tab:parameter}. The uncertainties listed in this table again derive from only the measurement uncertainties in the column densities. Some components do not have metallicity estimates because the \hi\ column density is not measured. 

\Cref{tab:parameter} shows a mixture of metallicity results. In particular, the metallicities in the high-velocity lines of J1008+3623 at $\textrm{v}\sim$ $-3100$ to $-2500$ \kms\ are roughly solar, while the low-speed lines at $\textrm{v} \sim$ $-250$ to $-100$ \kms\ are roughly a tenth solar. In J1225+4831, the metallicities range from roughly solar to a third solar. Q0249$-$222 and J1307+1230 both have substantially sub-solar metallicities. 

\subsection{Infall Candidates}

Several of the AAL complexes in our study have absorption-line components at small velocity shifts or positive velocities in the quasar frame that make them candidates for infall from the IGM. Most notable among these are the weak components 12 to 18 in J1008+3623, which have measured positive (infall) velocities from $\textrm{v}\sim$ 350 \kms\ to $\sim$ 650 \kms\ with estimated uncertainties of $\pm$100 \kms\ (Section 2.3). 

The ambiguous cases with very small negative velocity shifts, roughly from $-$500 to 0 \kms\, are components 9 to 11 in J1008+3623, components 5 to 8 in J1225+4831, components 4 to 7 in J1307+1230, and component 10 in J1633+1411. The uncertainties of their velocity shifts are roughly from 100 \kms\ to 220 \kms\ (Section 2.3). 

\subsection{Line-Lock Candidates}

Line-locked absorption-line systems refer to distinct systems that have velocity separations equal to the separation of a prominent doublet such as the \civ\ \lam 1548, 1551. The lines become locked at the doublet separation due to shadowing effects in radiatively-driven outflow. They can therefore be a signature of both the intrinsic nature of the absorption-line complex and radiative acceleration in the outflow \citep{Milne26, Scargle73, Braun89}. They might also occur in observed spectra due to chance alignments of unrelated absorption-line systems. However, observations of multiple line-locks in the same spectrum argue strongly for the reality of physical line locks in some cases \citep[e.g.,][]{Ganguly08, Hamann11}. Also, the large statistical study by \cite{Bowler14} showed that roughly two-thirds of SDSS quasars with multiple \civ\ NALs at speeds up to $\sim$12,000 \kms\ have at least one line-lock pair. Those results indicate that physical line-locking due to radiative forces is both real and common in quasar outflows. 

We consider two systems are locked when their velocity difference is a close match to the laboratory \civ\ doublet separation of 498 \kms. We find apparent line-locks between \civ\ doublets in components 1 and 6 of Q0249$-$222 and components 3 and 5 of J1225+4831. For Q0249$-$222, the measured separation between components 1 and 6 is $498\pm2$ \kms. The uncertainty in this velocity is remarkably small compared to the line widths, FWHM $\sim$ 30 \kms, and negligible compared to the gas velocities measured from the quasar systemic, roughly $-7640$ and $-$7140 \kms. The velocity separation between components 3 and 5 in J1225+4831 is $503\pm9$ \kms\ in lines with FWHM $\sim$ 40 \kms.

\subsection{Spatial Structure and Cloud Survival}

Partial covering of the quasar emission sources implies that the absorbing clouds are not much larger (and probably smaller) than the projected area of the emitting regions. We find evidence for partial covering in two of the quasars in our study, J1008+3623 (\Cref{tab:J1008}, \Cref{fig:J1008}) and J1633+1411 (\Cref{tab:J1633}, \Cref{fig:J1633} and Section 3.2). The high-speed ($\sim$2500$-$3100 \kms) and low-speed ($\sim$100$-$250 \kms) blueshifted lines in J1008+3623, and the high-speed ($\sim$5100$-$8200 \kms) blueshifted lines of J1633+1411 exhibit partial covering in \civ\ and \nv. These absorption lines sit atop strong broad emission lines (which are evident in SDSS spectra not plotted here), such that the partial covering could apply to the broad emission line regions (BLRs), which have a size roughly $0.1\sim1$ pc across in luminous quasars like our study \citep{Peterson93, Kaspi00}. Therefore these partial-covering absorbers have transverse sizes $\lesssim$1 pc. We also find evidence for partial covering in \ovi\ in J1008+3623, which do \textit{not} sit on strong emission lines. This absorber must partially cover the much smaller continuum source, such that the absorber sizes are conservatively $\lesssim$0.01 pc across \citep[][Hamann et al., in prep.]{Netzer92, Hamann11}. 

These small absorption-line clouds will have short survival times against dissipation if there is no external confinement \citep[e.g., magnetic disk winds,][]{deKool95, Proga03, Everett05}. The lifetime is of order the sound-crossing time, which can be approximated by 
\begin{equation}
t_{sc}=\frac{l}{c_s},
\end{equation}
where $c_s$ is the sound speed in an ideal monatomic gas, and $l$ is the characteristic cloud size \citep{Schaye01, Hamann01, Finn14}. For a nominal gas temperature of $10^4$ K in photoionised clouds, and $l \lesssim 0.01$ pc or $l \lesssim 1$ pc, the cloud survival times are $\lesssim$700 yr or $\lesssim$70,000 yr, respectively. The sound-crossing time is an upper limit to the dissipation time. If the internal cloud velocities are dominated by turbulence similar to the measured $b$ values (instead of thermal speeds), then the survival times would be even smaller. These short survival times suggest that the clouds are located near the quasars, where the flow times are shorter than the survival times, or they created in situ if the distances are larger (see C18 and Section 5.4 below). For example, outflow clouds moving at 3000 \kms\ will travel $\lesssim$200 pc in a survival time of $\lesssim$70,000 yr or only $\lesssim$2 pc in $\lesssim$700 yr. Small partial-covering outflow clouds at larger distances require either some type of confinement or a mechanism to create them far from the quasars. 

\section{Discussion}

The results presented in Sections 3 and 4 provide valuable constraints on the nature and origins of rich AAL complexes in luminous high-redshift quasars. We define AAL complexes for this study based on $\gtrsim$6 \civ\ absorption-line components spread over a velocity range $\lesssim$ 3000 \kms\ at velocity shifts $\lesssim$8000 \kms\ from the quasar emission-line redshift. The possible origins of AAL complexes in quasar environments include, (1) highly-structured clumpy outflows from the quasars, (2) interstellar clouds in the host galaxies that were (perhaps) shredded and dispersed into small clumps by high-speed quasar-driven outflows, and (3) condensations of cold-mode accreting gas from the IGM (see Section 1). We discuss each of these possibilities below, and provide a brief overview of the measured physical properties of the different AAL complexes in our study. 

The five quasars in our sample have \civ\ AAL complexes with from 7 up to 18 AAL components. The lines appear at velocity shifts from roughly $-8200$ to +640 \kms\ with Doppler $b$ values in \civ\ ranging from $\sim$5 to $\sim$100 \kms\ (\Cref{tab:Q0249,tab:J1008,tab:J1225,tab:J1307,tab:J1633}). The metallicities in well-measured systems range from roughly solar to $\sim$1/30 solar with total column densities from $\log N_{\rm{H}} ({\textrm{\cmN}}) \sim 18$ to 20 (Sections 4.1.2 and 4.1.1). All of the complexes have high degrees in ionisation based on \civ\ absorption, as required for inclusion in our study. Two out of the five complexes have at least one component with measured ionisations up to \ovi\ even though our ability to detect this ion is severely limited by blending in the \lya\ forest. None of them exhibit low-ionisation species such as \siii, \cii, or \mgii\ (Section 4.1.1). 

Only one quasar has AALs at positive velocity shifts, but they appear to be excellent candidates for infall. These are components 12 to 18 in J1008+3623 at v$\,\sim$ 350 to $\sim$ 640 \kms\ (Section 4.2). The other AAL components in this quasar have negative velocities consistent with outflows at speeds up to $-$3108 \kms. At least eight of the outflow components in J1008+3623, as well as seven components at higher outflow speeds up to $-$8224 \kms\ in J1633+1411, exhibit partial covering of the BLR and/or the continuum source, indicating that the absorbers have projected sizes (in the plane of the sky) $\lesssim1$ or $\lesssim0.01$ pc across, respectively. There is no partial covering in any of the other quasars/complexes (Sections 3.2 and 4.4). We find two apparent line-lock candidates in Q0249$-$222 and J1225+4831, which are likely signatures of outflows radiatively driven from the quasars (Section 4.3). 

\subsection{Intrinsic vs. Intervening}

It is known from statistical studies of large quasar samples that most AALs are intrinsic to the quasars, i.e., they form broadly within the environment of the quasar or its host galaxy (see Section 1 and refs. therein). Strong AAL \textit{complexes} are even more likely to be intrinsic to the quasars because their velocity ranges are not consistent with individual intervening galaxies and most (excepting only Q0249$-$222) have velocity ranges $>$1000 \kms\ too large even for rare massive galaxy clusters \citep[e.g.,][]{Koester07, Becker07, Munari13, Ruel14, Sohn17}. Previous studies of rich \civ\ complexes consider a variety of possible origins including outflows from the quasars/host galaxies \citep[e.g.,][]{Morris86, Dobrzycki99, Ganguly03, Misawa03, Misawa05, Misawa07, Simon10b, Hamann11}, gas condensations in cold-mode accreting gas in the quasar environments (see examples in C18), and intervening gas complexes unrelated to the quasar \citep{Morris86, Ganguly01b, Richards01a, Misawa03}. Most of these studies favor an intrinsic origin or at least note the difficulties described above for intervening interpretations. 

We favor an intrinsic origin for all five AAL complexes in our study because that provides a natural explanation for their clustered appearance in the spectra and most contain individual AAL systems with direct evidence for intrinsic origins. In particular, we find evidence for line locking and/or partial covering in the complexes in Q0249$-$222, J1008+3623, J1225+4831, and J1633+1411 (\Cref{tab:Q0249,tab:J1008,tab:J1225,tab:J1633,fig:Q0249,fig:J1008,fig:J1225,fig:J1633}). The fifth quasar, J1307+1230, does not provide direct evidence for an intrinsic origin. In fact, it appears to have sub-solar metallicities and it does not have measurable \nv\ absorption, which tends to be strong like \civ\ in known intrinsic systems \citep[e.g.,][]{Weymann81, Hartquist82, Hamann97c, Kuraszkiewicz02, Fox08, Perrotta18}. However, if we assume the AAL complex in J1307+1230 is due to intervening galaxies/halos, then it would require an odd coincidental distribution of seven galaxies with a velocity spread of $\sim$2000 \kms\ near the quasar redshift. The non-detection of the [\oiii] emission line in this quasar and others in our sample (Section 2) could be another clue that the AAL complex in J1307+1230 is intrinsic because weak [\oiii] could be caused by attenuation of the quasar's ionising radiation by outflowing AAL gas, similar to what has been proposed to explain weak [\oiii] lines in BAL quasars (see Section 5.3).

\subsection{Infall vs. Outflow}

The majority of absorption-line components in the AAL complexes we study are at large negative velocities, v$\, < -$1000 \kms\ up to roughly $-8200$ \kms. This far exceeds the estimated redshift uncertainties (Section 2.3). Given the evidence for the intrinsic origins of these complexes in Section 5.1, we attribute all of these blueshifted systems to high-speed quasar-driven outflows. It is important to note that they are \textit{quasar-driven} outflows because the alternative, e.g., starburst-driven winds in the host galaxies, are known to have smaller velocities from typically $\sim$100 to $\sim$1000 \kms\ (e.g., \citealt{Heckman00}, see also \citealt{Misawa07, Nestor08, Wild08, Simon10} for more discussion). 

Four out of the five AAL complexes in our study also have systems at small negative velocities v$\,\gtrsim$ $-$540 \kms, e.g., components 9 to 11 in J1008+3623, 5 to 8 in J1225+4831, 4 to 7 in J1307+1230, and component 10 in J1633+1411. The outflow nature of these systems is ambiguous given the estimated redshift uncertainties of $\sim$100 \kms\ up to 221 \kms\ in J1307+1230 (Section 2.3). They are consistent with being at rest with respect to the quasars and their host galaxies or even involved in low-speed infall. However, if we take the small measured negative velocities at face value, they are consistent with starburst-driven winds, low-speed quasar outflows, or, perhaps, formerly high-speed quasar outflows that were slowed down as they became mass loaded with ambient interstellar gas in the host galaxies. 

The only lines at positive velocities in our study are components 12 to 18 in J1008+3623. These systems are excellent candidates for infall toward the quasar because their measured velocity shifts from $\sim$350 to $\sim$650 \kms\ are significantly larger than the estimated uncertainties ($\pm100$ kms). We discuss these infall candidates further in Section 5.4. 

In addition to velocity shifts, the gas metallicities might also provide clues to the infall/outflow nature of the AALs. For example, outflow systems ejected from the quasar/galactic nuclear environments could naturally be metal rich, while infall systems from the distant IGM could be generally metal poor \citep{Hamann99, Arav01, Hamann02, Dietrich03, Warner04, Gabel05b, Nagao06, Simon10}. However, there are caveats to this simple interpretation. Outflow systems might also be metal poor if they are mixed with interstellar gas swept by the quasar outflow in the outer host galaxies. Infall systems could be (moderately) metal rich if the gas is part of a fountain-like circulation pattern, e.g., previously ejected from the quasar/host galaxy while remaining gravitationally bound so we observe it now falling back \citep[e.g.,][]{Sancisi08, Marinacci11, Wada12, Wada15, Wada16, Tumlinson17}. 

In our study, the roughly solar metallicities we find in the components 1 to 8 of J1008+3623 (\Cref{tab:parameter}) are consistent with the naive expectations for a quasar-driven outflow, while the sub-solar metallicities in all components of Q0249$-$222 and the component 1 of J1307+1230 (\Cref{tab:parameter}) might indicate that they are mixed with ambient gas at large radii in the host galaxies. Unfortunately, we do not have firm metallicity constraints for the candidate infall systems in J1008+3623 (components 12-18) because we do not detect the \hi\ Lyman lines (\Cref{tab:J1008}). However, the lower limits $\gtrsim$0.03 times solar (\Cref{tab:parameter}) rule out primordial gas and direct infall from the distant IGM. They may require some significant enrichment inside a galaxy, perhaps in a galactic fountain as noted above. 

\subsection{[\oiii] Emission}

We observed J1008+3623 and J1307+1230 in the K band to measure redshifts from their [\oiii] or \hb\ emission lines (Section 2). It is surprising that we did not detect these lines in our spectra. Weak [\oiii] and \hb\ lines might be related to the quasar luminosities in the Baldwin effect \citep{Baldwin77}, which describes a trend for decreasing emission-line rest equivalent width (REW) with increasing quasar continuum luminosities \citep[also][]{Brotherton96, Zhang11, Stern12a, Stern12b, Shen16b}. J1008+3623 and J1307+1230 have similar high bolometric luminosities of $L\sim 3\times10^{47}$ erg s$^{-1}$ \citep{Krawczyk13}. \cite{Shen16b} measured [\oiii] line strengths in a sample of 74 quasars with high luminosities, $L\sim 10^{46.2}$ to $\sim$$10^{48.2}$ erg s$^{-1}$, and high redshifts, $z\sim$1.5 to $\sim$3.5, broadly similar to J1008+3623 and J1307+1230. They found that the quasars in their sample show weaker [\oiii] emission lines than typical SDSS quasars at lower luminosities (and $z < 1$). Thus the Baldwin Effect does appear to play some role. However, \citet{Shen16b} detected [\oiii] in all but one out of their 74 quasars, and they measured a median [\oiii] equivalent width of REW([\oiii])~$\sim 13$ \AA\ and a median full width at half maximum FWHM([\oiii])~$\sim 1000$ \kms . If we assume the quasars in our study have similar [\oiii] profiles, then the minimum REW([\oiii]) we could detect in our near-IR spectra is $\sim$3.3 \AA. We conclude that the [\oiii] emission lines in J1008+3623 and J1307+1230 are exceptionally weak compared to similar luminous quasars at these redshifts. 

A possible explanation for this might come from noting that [\oiii] emission lines are also characteristically weak in BAL quasars, e.g., by a factor of $\sim$2 compared to non-BAL quasars at similar redshifts and luminosities \citep{Yuan03, Wills04}. In particular, they measured a median REW([\oiii])~$\lesssim 3$ \AA\ for BAL quasars, which is less than the minimum REW([\oiii]) we could detect in our near-IR spectra, and consistent with our non-detection result. The origins of this behavior are not understood yet, but some previous studies attribute weaker [\oiii] emission in BAL quasars to attenuation of the quasar's ionising radiation by the BAL outflow gas before it reaches the [\oiii]-emitting regions much farther out \citep{Boroson92, Turnshek97, Yuan03}. Regardless of the physical cause, the absence of [\oiii] emission in two out of two quasars with rich AAL complexes that we measured in the near-IR, J1008+3623 and J1307+1230, suggests that there is some relationship between these AAL complexes and BAL outflows. One obvious relationship is that BALs and the AAL complexes we study all form in outflows. More work is needed to determine if the outflows or other material connected to rich AAL complexes can provide sufficient far-UV attenuation to affect the [\oiii] line strengths. It is worth noting, however, that this could be a natural consequence of AAL complexes forming at an interface where high-speed quasar outflows shred and disperse interstellar clouds (Section 1).

\subsection{J1008+3623}

The AAL complex in J1008+3623 is an interesting case that merits more discussion. The lines in this complex can be divided into three velocity groups that might identify three different physical environments along our line of sight (\Cref{tab:J1008,fig:J1008}). The first group involves components 1 to 8 at measured velocity shifts $-3100\lesssim v\lesssim-1250$ \kms\ and Doppler widths $20\lesssim b\lesssim56$ \kms. Several of these components exhibit partial covering indicative of intrinsic absorption. The metallicities in this line group are roughly solar or higher (\Cref{tab:parameter}), consistent with a quasar-driven outflow ejected from the metal-rich environs of a galactic nucleus \citep{Hamann99, Arav01, Hamann02, Dietrich03, Warner04, Gabel05b, Nagao06, Simon10}. 

The second velocity group in J1008+3623 involves components 9 to 11 at $-250\lesssim v\lesssim-100$ \kms\ with narrower line profiles than the outflow group, $18\lesssim b\lesssim24$ \kms. All of these components exhibit partial covering and two of the three have metallicities that are roughly one dex below solar. In Section 4.4, we argued that the small cloud sizes required by partial covering imply that the absorbing clouds are either close to the quasar or created in situ at larger distances in the host galaxy. The very low velocities in this second line group indicate further that the clouds 1) can travel negligibly small distances during their short survival times, and 2) are outside the gravitational sphere of influence of the central black hole. We conclude that they are almost certainly created in situ in the extended host galaxy. This interpretation is supported by the sub-solar metallicities, which are consistent with ambient interstellar gas in the outer regions of the galaxy.

It is interesting to consider that velocity groups 1 and 2 in J1008+3623 might be physically related. In particular, they might identify an environment where a high-speed quasar-driven outflow is interacting with ambient interstellar material in the host galaxy. These interactions are predicted by quasar feedback models to shred and disperse interstellar clouds in the galaxies, thus creating small cloud fragments at a range of velocities \citep{Hopkins10, Faucher11, Faucher12b}. This situation could produce highly-structured AAL complexes like we observe in J1008+3623 and other quasars in this study \citep[see also C18;][and refs. in Section 1]{Simon10, Simon12}. It provides a natural explanation for the in-situ creation of small partial-covering clouds far from the quasars. The cloud fragments more closely related to the original outflow could have higher velocities, larger velocity dispersions, and higher metallicities (as in velocity group 1 in J1008+3623), while those more closely tied to the shredded interstellar clouds would have lower speeds, smaller $b$ values, and lower metallicities (group 2). 

The third velocity group in this AAL complex is components 12 to 18 at positive velocities $350\lesssim v\lesssim650$ \kms\ and small velocity dispersions, $5\lesssim b\lesssim30$ \kms. As noted in Section 5.2, these AALs are the best candidates for infall in our study. Their positive velocities and narrow profiles are consistent with formation in condensations in cold-mode accreting gas from the intergalactic or circumgalactic medium \citep{Keres09, Keres12, Fumagalli11, Hafen16}. However, their moderate metallicities (with lower limits of $\sim$0.03 times solar, \Cref{tab:parameter}) appear to require significant enrichment inside the galaxy. Therefore, the infall we measure could be part of a fountain-like circulation pattern of gas that is gravitationally-bound to the host galaxy. 

It is likely that infall and outflow occur together generally if cold-mode accretion is, indeed, involved in fueling starbursts and quasar activity that both drive outflows and feedback in the host galaxies \citep{Costa14, Nelson15, Suresh15}. The three velocity groups in the AAL complex in J1008+3623, exhibiting outflow (group 1), infall (group 3), and \textit{possible} feedback effects (group 2), might provide the first direct measurements of these diverse processes occurring together around one quasar during an active and perhaps early stage of massive galaxy evolution at redshift $z\sim$3.13. 

\section{Summary}

We analyze rest-frame UV spectra of five quasars at redshifts $3.1< z<4.4$ that have rich multi-component complexes of \civ\ associated absorption lines (AALs). AALs provide important constraints on the radial gas kinematics, metallicities, and physical conditions in quasar environments during an active stage of high-redshift galaxy evolution. Rich AAL complexes are particularly interesting because they might uniquely measure a feedback phenomenon where ISM clouds in the host galaxies are being shredded and dispersed into small fragments by high-speed quasar outflows. Some AAL complexes (e.g., J1008+3623 in this study plus another in C18) include candidate infall lines that could form in  condensations in cold-mode accreting gas from the circumgalactic medium or, perhaps, in a fountain-like circulation pattern in the galaxies. The main results of our study are the following: 


1) All of the AAL complexes are highly ionised with measured lines of \civ\ by definition, but also \nv\ and \ovi\ in some cases. None of the complexes have lines of low ionisation ions such as \siii, \cii, or \mgii\ (Section 4.1.1). 

2) At least half of the AAL systems in two complexes (in J1008+3623 and J1633+1411) exhibit partial covering of the background emission source (Section 4.4). This implies that the absorbers are composed of small clouds with characteristic sizes $\lesssim1$ pc and possibly $\lesssim0.01$ pc (based on \ovi\ in J1008+3623 that partially cover the quasar continuum source). These tiny clouds will have short survival times of 70,000 yr or even 700 yr if they are not confined by an external pressure, suggesting that the clouds in these complexes are located very close to the central quasars or created in situ at larger distances, perhaps via the shredding of interstellar clouds in the host galaxies.

3) We argue that all five AAL complexes in our study are intrinsic to the quasar environments. There is direct evidence for intrinsic origins in four of the five quasars based on partial covering, line-locked \civ\ systems, and/or strong \nv\ absorption relative to \civ . We also note that the alternative explanation (that these rich absorption-line complexes form in unrelated intervening galaxies/halos near the quasar redshifts) is implausible (Section 5.1).

4) All five AAL complexes in our study have at least some lines at large negative velocities $v\lesssim -1000$ \kms\ up roughly to $-8200$ \kms . We attribute these highly-blueshifted systems to high-speed quasar-driven outflows. In our limited sample, these systems appear most likely to have line locks, partial covering, strong \nv , and solar metallicities (Sections 4.3, 4.4 and 4.1.2). They also tend to have broader profiles, with Doppler $b$ parameters in \civ\ reaching $\sim$100 \kms\ (Section 3). 

5) Four of the five AAL complexes we study have lines at small negative velocities, from $v\sim0$ up to roughly $-500$ \kms . These lines have ambiguous origins due to the redshift uncertainties. They might form in ambient interstellar/halo gas or in low-speed outflows from the quasars or galactic starbursts. Another intriguing possibility is that they identify a feedback phenomenon at an interface where higher-speed quasar-driven outflows are impacting and shredding interstellar clouds in the quasar host galaxies (Sections 5.2 and 5.4).

6) The AAL complex in one quasar, J1008+3623, has several absorption systems with significant positive velocities, $350\lesssim v\lesssim650$ \kms, and unusually narrow \civ\ line profiles. These systems are excellent candidates for infall toward the quasar. They might identify gas condensations associated with cold-mode accretion from the circumgalactic medium. However, the derived moderate metallicities, $\gtrsim$0.03 times solar, suggest that the gas was substantially enriched inside a galaxy. These infall lines might, therefore, form in a fountain-like circulation pattern of gas that is gravitationally bound to the quasar host galaxy (Section 5.4).

\begin{figure*}
\centering
\includegraphics[width=1.0\textwidth]{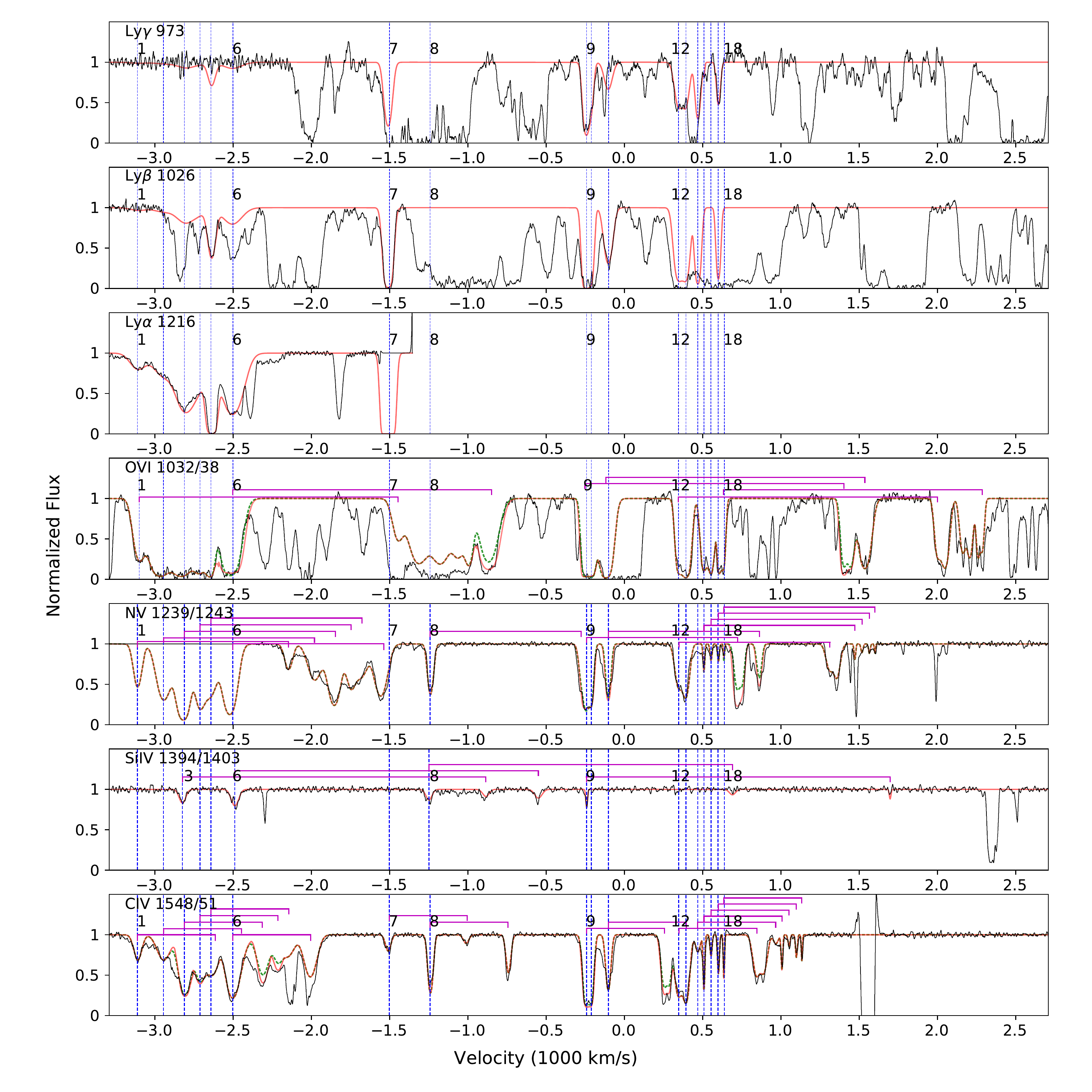}
\caption{J1008+3623: Normalized line profiles in the KeckI-HIRES spectra plotted on a velocity scale relative to the quasar redshift (\Cref{tab:sample}). The spectra are shown in black, and the final fitting lines are shown in red (grey solid). The blue dash lines are identified components from 1 to 18, and the brackets show the doublets. The velocities pertain to the short-wavelength lines in the doublets. The green dash lines show the predicted lines based on \civ\ \lam 1548, \nv\ \lam 1239, or \ovi\ \lam 1032 assuming $C_0=1$.\label{fig:J1008}}
\end{figure*}

7) The AAL complex in J1008+3623 is the only one in our study with lines in all three velocity groups described above: high-speed quasar-driven outflow lines that are broad and metal-rich (roughly solar), low-speed lines with intermediate metallicities ($\sim$0.1 times solar), and unusually narrow infall lines that are relatively metal poor at $\gtrsim$0.03 times solar (\Cref{fig:J1008,tab:J1008,tab:parameter}). Together the three line groups in this AAL complex might provide the first direct measurements of infall, quasar-driven outflows, and cloud-shredding feedback effects occurring simultaneously around one quasar during an active and (perhaps) early stage of high-redshift massive galaxy evolution (Section 5.4).

\clearpage

\clearpage

\clearpage

\clearpage

\clearpage

\clearpage

\clearpage

\clearpage

\clearpage
\startlongtable
\begin{deluxetable*}{lccccc}
\tablecaption{Quasar data including the name, emission-line redshift, telescope-instrument used for observations, observation date (dd/mm/year), rest wavelength ranges, and spectral resolution $R=\lambda /\Delta\lambda$.\label{tab:sample}}
\tabletypesize{\scriptsize}
\tablehead{
\colhead{Quasar} & \colhead{$z_{\rm{em}}$} & \colhead{Instrument} & \colhead{Obs. Date} & \colhead{$\lambda_{\rm{rest}}$ (\r{A})} & \colhead{R}
}
\startdata
Q0249$-$222 (UM 678)& $3.205\pm0.009^a$ ($\pm642$ \kms) & VLT-UVES & 09/22/2003 & 775-1075 & 80000\\
\vspace{2mm} \nodata  & \nodata &\nodata  & \nodata & 1130-1625 & 110000\\
\vspace{2mm}SDSS J100841.22+362319.3 & $3.1291\pm0.0014^b$ ($\pm100$ \kms) & Keck-HIRES & 03/26-27/2007 & 885-1965 & 45000\\
\vspace{2mm}SDSS J122518.66+483116.3 & $3.0960\pm0.0014^b$ ($\pm100$ \kms) & Keck-HIRES & 03/27/2007 & 900-1980 & 45000\\
SDSS J130710.25+123021.6 & $3.2104\pm0.0031^b$ ($\pm221$ \kms) & Magellan-MIKE & 03/20/2007 & 795-1225 & 57600\\
\vspace{2mm} \nodata &\nodata  & \nodata & \nodata & 1150-2240 & 45150\\
SDSS J163319.62+141142.0 & $4.3753\pm0.0018^b$ ($\pm100$ \kms) & Keck-HIRES & 06/27/2008 & 880-1640 & 45000\\
\enddata
\tablenotetext{a}{\citet{Tytler92}}
\tablenotetext{b}{\citet{Hewett10}}
\end{deluxetable*}

\clearpage
\startlongtable
\begin{deluxetable*}{cccccccc}
\tablecaption{Individual absorption lines of Q0249$-$222. Columns show component number, absorption redshift ($z_{abs}$) and the corresponding velocity shift ($\textrm{v}$) relative to the emission-line redshift, line identification and rest wavelength, observation wavelength, Doppler b parameter, logarithm of column density, covering fraction, and notes (sat=saturated line, bl=blended with neighboring systems or unrelated lines (e.g., lines in the Lyman forest), nm=no measurements due to a gap between echelle orders).\label{tab:Q0249}}
\tabletypesize{\scriptsize}
\tablehead{
\colhead{$\#$ } & \colhead{$z_{abs}$} & 
\colhead{ID} & \colhead{$\lambda_{obs}$} & 
\colhead{$b$} & \colhead{$\log N$} & 
\colhead{$C_0$ } & \colhead{Notes}\\ 
\colhead{} & \colhead{$\textrm{v}$ (\kms)} & \colhead{} & \colhead{(\AA)} &  \colhead{(\kms) } &
\colhead{(\cmN)} & \colhead{} & \colhead{} 
} 
\startdata
1 & 3.0993 & \lya\ 1216 & 4983.28 & $26.7\pm0.7$ & $13.19\pm0.01$ & 1.0$^c$ & \\
  & $-7636\pm642$ & \civ\ 1548 & 6346.52 & $23.5\pm 2.0$ & $13.24\pm0.03$ & 1.0$^b$ & \\
 &  & \civ\ 1551 & 6357.11 & --- & --- & --- & bl\\ 
 \hline
2 & 3.1013 & \lye\ 938 & --- & --- & --- & --- & bl \\
& $-7490\pm642$ & \lyd\ 950 & --- & --- & --- & --- & bl \\
  & & \lyc\ 973 & 3988.49 & $22.2\pm0.8$ & $14.76\pm0.02$ & 1.0$^{a,b}$ &  \\
& & \lyb\ 1026 & 4206.60 & --- & --- & --- & \\
& & \lya\ 1216 & 4985.52 & --- & --- & --- & sat \\
& & \siiv\ 1394 & 5715.90 & $8.4\pm 0.8$ & $<12.7$ & 1.0$^c$ & bl \\
&  & \siiv\ 1403 & 5752.88 & --- & --- & --- & nm\\ 
 & & \civ\ 1548 & 6349.56 & $19.8\pm 0.9$ & $13.67\pm0.02$ & 1.0$^b$ & \\
& & \civ\ 1551 & 6360.15 & --- & --- & --- & \\ 
\hline
3 & 3.1022 & \lye\ 938 & --- & --- & --- & --- & bl \\
 & $-7424\pm642$ & \lyd\ 950 & --- & --- & --- & --- & bl \\
  & & \lyc\ 973 & 3989.63 & $20.8\pm1.4$ & $14.42\pm0.09$ & 1.0$^{a,b}$ & bl \\
& & \lyb\ 1026 & --- & --- & --- & --- & bl\\
 & & \lya\ 1216 & --- & --- & --- & --- & sat \&\ bl \\
 & & \siiv\ 1394 & 5717.53 & $6.6\pm 1.8$ & $12.11\pm0.08$ & 1.0$^c$ & \\
&  & \siiv\ 1403 & 5754.49 & --- & ---& --- & nm \\ 
 & & \civ\ 1548 & 6351.07 & $31.8\pm 2.4$ & $13.84\pm0.03$ & 1.0$^b$ & \\
&  & \civ\ 1551 &  6361.66 & --- & --- & --- & \\ 
\hline
4 & 3.1029 & \lye\ 938 & --- & --- & --- & --- & bl \\
 & $-7373\pm642$ & \lyd\ 950 & 3896.73 & $21.5\pm1.0$ & $15.04\pm0.02$ & 1.0$^{a,b}$ &  \\
 & & \lyc\ 973 & 3990.25 & --- & --- & --- &  \\
& & \lyb\ 1026 & --- & --- & --- & --- & sat \&\ bl\\
 & & \lya\ 1216 & --- & --- & --- & --- & sat \&\ bl \\
 & & \siiv\ 1394 & 5718.48 & $5.7\pm 0.4$ & $12.52\pm0.02$ & 1.0$^c$ & \\
&  & \siiv\ 1403 & 5755.44 & ---&--- & --- & nm\\ 
 & & \civ\ 1548 & 6352.15 & $17.0\pm 0.7$ & $13.77\pm0.03$ & 1.0$^b$ & \\
&  & \civ\ 1551 &  6362.74 & ---& ---& --- & \\ 
\hline
5 & 3.1049 & \lye\ 938 & --- & --- & --- & --- & bl \\
 & $-7226\pm642$ & \lyd\ 950 & --- & --- & --- & --- & bl \\
 & & \lyc\ 973 & --- & --- & --- & --- & bl \\
& & \lyb\ 1026 & 4210.31 & $43.3\pm1.9$ & $14.69\pm0.04$ & 1.0$^{a,b}$ &  \\
& & \lya\ 1216 & 4989.86 & --- & --- & --- & sat \\
 & & \civ\ 1548 & 6355.23 & $35.1\pm 1.8$ & $13.54\pm0.02$ & 1.0$^b$ & \\
&  & \civ\ 1551 &  6365.83 & ---& ---& --- & \\ 
\hline
6 & 3.1061 & \lye\ 938 & --- & --- & --- & --- & bl \\
 & $-7138\pm642$  & \lyd\ 950 & --- & --- & --- & --- & bl \\
 & & \lyc\ 973 & --- & --- & --- & --- & bl \\
& & \lyb\ 1026 & 4211.74 & $22.3\pm0.8$ & $14.37\pm0.02$ & 1.0$^{a,b}$ & \\
&  & \lya\ 1216 & 4991.56 & --- & --- & --- & sat \\
 & & \siiv\ 1394 & 5722.76 & $4.9\pm 0.6$ & $11.94\pm0.04$ & 1.0$^c$ & \\
&  & \siiv\ 1403 & 5759.75 & ---& ---& ---& nm\\ 
 & & \civ\ 1548 & 6357.05 & $17.6\pm 0.7$ & $13.65\pm0.02$ & 1.0$^b$ & bl \\
& & \civ\ 1551 &  6367.66 & ---& ---& --- & \\ 
\hline
7 & 3.1071 & \lye\ 938 & --- & --- & --- & --- & bl \\
 & $-7066\pm642$  & \lyd\ 950 & --- & --- & --- & --- & bl \\
 & & \lyc\ 973 & --- & --- & --- & --- & bl \\
& & \lyb\ 1026 & 4212.76 & $15.8\pm0.5$ & $14.37\pm0.02$ & 1.0$^{a,b}$ &  \\
&  & \lya\ 1216 & 4992.75 & --- & --- & --- & sat \&\ bl \\
 & & \siiv\ 1394 & 5724.28 & $5.4\pm 0.9$ & $12.16\pm0.04$ & 1.0$^c$ & \\
&  & \siiv\ 1403 & 5761.28 & --- & ---& --- & nm\\ 
& & \civ\ 1548 & 6358.58 & $15.8\pm 2.4$ & $13.16\pm0.04$ & 1.0$^b$ & \\
&  & \civ\ 1551 & 6369.18 & --- &--- & --- & \\ 
\hline
8 & 3.1075 & \lya\ 1216 & 4993.27 & 11 & $12.5\pm0.2$ & 1.0$^c$ & bl, adopt $b$ from \civ\ \\
& $-7037\pm642$ & \civ\ 1548 & 6359.24 & $10.8\pm 1.8$ & $12.84\pm0.08$ & 1.0$^b$ & \\
&  & \civ\ 1551 & 6369.85 & ---& ---&--- & \\ 
\enddata
\tablenotetext{a}{Heavily saturated lines, whose $C_0$ equals the observed depth of the line.}
\tablenotetext{b}{Unsaturated doublets, where we solve for $C_0$ from fitting.}
\tablenotetext{c}{Unsaturated single lines, where we adopt $C_0=1$ because it is not constrained.}
\end{deluxetable*}

\clearpage
\startlongtable
\begin{deluxetable*}{cccccccc}
\tablecaption{Individual absorption lines of J1008+3623. See \Cref{tab:Q0249} for descriptions of the table contents. For the notes, sat=saturated line, bl=blended with neighboring systems or unrelated lines (e.g., lines in the Lyman forest), nm=no measurements.\label{tab:J1008}}
\tabletypesize{\scriptsize}
\tablehead{
\colhead{$\#$ } & \colhead{$z_{abs}$} & 
\colhead{ID} & \colhead{$\lambda_{obs}$} & 
\colhead{$b$} & \colhead{$\log N$} & 
\colhead{$C_0$ } & \colhead{Notes}\\ 
\colhead{} & \colhead{$\textrm{v}$ (\kms)} & \colhead{} & \colhead{(\AA)} &  \colhead{(\kms) } &
\colhead{(\cmN)} & \colhead{} & \colhead{} 
} 
\startdata
1 & 3.0863 & \lya\ 1216 & 4970.33 & $66.3\pm2.3$ & $13.28\pm0.01$ & 1.0$^c$ & \\
& $-3108\pm102$ & \ovi\ 1032 & 4216.73 & $44.7\pm1.6$ & $14.49\pm0.01$ & 1.0$^c$ & \\
& & \ovi\ 1038 & 4240.03 & --- & --- & --- & bl with \lyb\ \\
& & \nv\ 1239 & --- & --- & --- & --- & nm\\
& & \nv\ 1243 & 5078.45 & $30.1\pm3.1$ & $13.94\pm0.06$ & 1.0$^c$ &  \\
& & \civ\ 1548 & 6326.41 & $33.3\pm1.2$ & $13.46\pm0.01$ & 1.0$^c$ & \\
& & \civ\ 1551 & 6336.96 & --- &--- & ---& bl \\
\hline
2 & 3.0886 & \lya\ 1216 & 4970.33 & 56 & $13.5\pm0.1$ & 1.0$^c$ & bl, adopt $b$ from \civ\ \\
& $-2942\pm102$ & \ovi\ 1032 & --- & --- & --- & --- & bl, sat \\
& & \ovi\ 1038 & --- & --- & --- & --- & bl with \lyb\ \\
& & \nv\ 1239 & --- & --- & --- & --- & nm\\
& & \nv\ 1243 & 5081.31 & 56 & $14.4\pm0.1$ & 1.0$^c$ & bl, adopt $b$ from \civ\ \\
& & \civ\ 1548 & 6329.97 & $55.9\pm5.3$ & $13.68\pm0.04$ & 1.0$^c$ & \\
& & \civ\ 1551 & 6340.50 & --- &--- &--- & bl\\
\hline
3 & 3.0904 & \lyb\ 1026 & 4195.61 & --- & --- & --- & bl\\
& $-2807\pm102$ & \lya\ 1216 & 4972.58 & $72.2\pm2.1$ & $14.05\pm0.01$ & 1.0$^c$ & bl\\
& & \ovi\ 1032 & --- & --- & --- & --- & bl, sat \\
& & \ovi\ 1038 & --- & --- & --- & --- & bl with \lyb\ \\
& & \nv\ 1239 & --- & --- & --- & --- & nm\\
& & \nv\ 1243 & 5083.55 & $39.8\pm2.3$ & $14.67\pm0.05$ & 1.0$^c$ & \\
& & \siiv\ 1394 & 5700.73 & $28.0\pm6.1$ & $12.68\pm0.08$ & 1.0$^d$ & \\
& & \siiv\ 1403 & 5737.58 & --- & --- & --- & \\
& & \civ\ 1548 & 6332.76 & $40.3\pm1.0$ & $14.48\pm0.04$ & $0.83\pm0.04^b$ & \\
& & \civ\ 1551 & 6343.29 & --- & --- & --- & \\
\hline
4 & 3.0917 & \lyb\ 1026 & --- & --- & --- & --- & bl\\
& $-2712\pm102$ & \lya\ 1216 & --- & 37 & $<13.70$ & 1.0$^c$ & bl, adopt $b$ from \civ\ \\
& & \ovi\ 1032 & --- & --- & --- & --- & bl, sat \\
& & \ovi\ 1038 & --- & --- & --- & --- & bl with \lyb\ \\
& & \nv\ 1239 & --- & --- & --- & --- & nm\\
& & \nv\ 1243 & 5085.16 & 37 & $14.25\pm0.02$ & 1.0$^c$ & bl, adopt $b$ from \civ\  \\
& & \civ\ 1548 & 6334.77 & $37.0\pm0.9$ & $14.23\pm0.12$ & $0.71\pm0.05^b$ & \\
& & \civ\ 1551 & 6345.31 & --- & --- & --- & \\
\hline
5 & 3.0926 & \lyb\ 1026 & 4197.86 & $30.7\pm0.7$ & $14.38\pm0.01$ & 1.0$^b$ & \\
& $-2646\pm102$ & \lya\ 1216 & 4975.25 & --- & --- & --- & \\
& & \ovi\ 1032 & --- & --- & --- & --- & bl, sat\\
& & \ovi\ 1038 & --- & --- & --- & --- & bl\\
& & \nv\ 1239 & --- & --- & --- & --- & nm\\
& & \nv\ 1243 & 5086.28 & 53 & $14.28\pm0.02$ & 1.0$^c$ & bl, adopt $b$ from \civ\ \\
& & \civ\ 1548 & 6336.16 & $52.6\pm1.5$ & $\geq13.86$ & $\leq1.0^b$ & \\
& & \civ\ 1551 & 6346.71 & --- & --- & --- & bl \\
\hline
6 & 3.0946 & \lyb\ 1026 & 4199.91 & --- &--- & --- & bl\\
& $-2503\pm102$ & \lya\ 1216 & 4977.68 & $80.3\pm2.5$ & $<14.2$ & 1.0$^c$ & bl with \siiii\ \\
& & \ovi\ 1032 & 4225.30 & $66.6\pm1.2$ & $15.29\pm0.08$ & $0.91\pm0.05^b$ & \\
& & \ovi\ 1038 & 4248.64 & --- & --- & --- & \\
& & \nv\ 1239 & --- & --- & --- & --- & nm\\
& & \nv\ 1243 & 5088.77 & $55.6\pm1.2$ & $14.56\pm0.01$ & 1.0$^c$ & \\
& & \siiv\ 1394 & 5707.15 & $32.0\pm5.2$ & $12.83\pm0.06$ & 1.0$^d$ & \\
& & \siiv\ 1403 & 5744.05 & ---&--- & ---&\\
& & \civ\ 1548 & 6339.26 & $51.0\pm0.9$ & $\geq14.23$ & $\leq1.0^b$ & bl\\
& & \civ\ 1551 & 6349.81 &--- &--- &--- & bl \\
\hline
7 & 3.1083 & \lyc\ 973 & --- & --- & --- & --- & bl\\
& $-1509\pm102$ & \lyb\ 1026 & 4213.97 & $29.6\pm0.8$ & $15.04\pm0.03$ & 1.0$^b$ & \\
& & \lya\ 1216 & --- & --- & --- & --- & nm \\
& & \civ\ 1548 & 6360.47 & $22.4\pm4.0$ & $13.08\pm0.07$ & 1.0$^b$ & \\
& & \civ\ 1551 & 6371.05 & --- & --- &--- & \\
\hline
8 & 3.1120 & \lyc\ 973 & --- & --- & --- & --- & bl\\
& $-1238\pm102$ & \lyb\ 1026 & --- & --- & --- & --- & bl with \ovi\ \\
& & \lya\ 1216 & --- & --- & --- & --- & nm \\
& & \nv\ 1239 & 5094.03 & $22.1\pm1.5$ & $13.89\pm0.03$ & 1.0$^c$ & \\
& & \nv\ 1243 & 5110.39 & --- & --- & --- & bl\\
& & \siiv\ 1394 & 5730.91 & $25.5\pm7.4$ & $12.54\pm0.09$ & 1.0$^d$ & \\
& & \siiv\ 1403 & 5767.96 & --- & --- & --- & \\
& & \civ\ 1548 & 6366.20 & $19.0\pm0.9$ & $13.76\pm0.02$ & 1.0$^b$ & \\
& & \civ\ 1551 & 6376.79 & ---& ---& ---& \\
\hline
9 & 3.1257 & \lyc\ 973 & 4014.31 & 24 & $15.09\pm0.13$ & 1.0$^b$ & bl, adopt $b$ from \civ\ \\
& $-243\pm102$ & \lyb\ 1026 & 4231.81 & --- & --- & --- & ---\\
& & \lya\ 1216 & --- & --- & --- & --- & nm \\
& & \ovi\ 1032 & 4257.39 & 24 & $14.92\pm0.10$ & $0.97\pm0.02^b$ & bl, adopt $b$ from \civ\ \\
& & \ovi\ 1038 & 4280.91 & --- & --- & --- & ---\\
& & \nv\ 1239 & 5111.00 & 24 & $14.73\pm0.07$ & $0.80\pm0.03^b$ & bl, adopt $b$ from \civ\ \\
& & \nv\ 1243 & 5127.42 & --- & --- & --- & ---\\
& & \siiv\ 1394 & 5750.22 & $6.9\pm2.2$ & $12.23\pm0.10$ & 1.0$^d$ & \\
& & \siiv\ 1403 & 5791.52 & --- & --- & --- & \\
& & \civ\ 1548 & 6387.41 & $23.6\pm0.4$ & $14.20\pm0.02$ & $0.92\pm0.02^b$ & bl\\
& & \civ\ 1551 & 6398.04 & ---& ---& --- & ---\\
\hline
10 & 3.1261 & \lyc\ 973 & 4014.71 & 18 & $14.15\pm0.20$ & 1.0$^b$ & bl, adopt $b$ from \civ\ \\
& $-210\pm102$ & \lyb\ 1026 & 4232.22 & --- & --- & --- & bl\\
& & \lya\ 1216 & --- & --- & --- & --- & nm \\
& & \ovi\ 1032 & 4257.81 & 18 & $14.76\pm0.10$ & $0.97\pm0.02^b$ & bl, adopt $b$ from \civ\ \\
& & \ovi\ 1038 & 4281.32 & --- & ---& ---& ---\\
& & \nv\ 1239 & 5111.50 & 18 & $14.16\pm0.07$ & $0.80\pm0.03^b$ & bl, adopt $b$ from \civ\ \\
& & \nv\ 1243 & 5127.92 & ---& ---& ---& ---\\
& & \civ\ 1548 & 6388.03 & $17.7\pm0.6$ & $14.00\pm0.04$ & $0.92\pm0.02^b$ & bl\\
& & \civ\ 1551 & 6398.66 & --- & ---& ---& ---\\
\hline
11 & 3.1276 & \lyc\ 973 & --- & --- & --- & --- & bl\\
& $-102\pm102$ & \lyb\ 1026 & 4233.76 & $33.4\pm0.8$ & $14.51\pm0.04$ & 1.0$^c$ & \\
& & \lya\ 1216 & --- & --- & --- & --- & nm \\
& & \ovi\ 1032 & 4259.35 & $43.6\pm1.3$ & $14.93\pm0.02$ & $1.0^b$ & bl\\
& & \ovi\ 1038 & 4282.88 &--- & --- & ---& \\
& & \nv\ 1239 & 5113.35 & $20.0\pm0.9$ & $14.28\pm0.04$ & $0.74\pm0.04^b$ & \\
& & \nv\ 1243 & 5129.78 & ---& ---& ---& \\
& & \civ\ 1548 & 6390.35 & $19.5\pm0.7$ & $13.71\pm0.01$ & 1.0$^c$ & \\
& & \civ\ 1551 & 6400.98 & --- &--- & ---& bl\\
\hline
12 & 3.1338 & \lyc\ 973 & 4022.19 & 30 & $<14.9$ & 1.0$^c$ & bl, adopt $b$ from \civ\ \\
& $346\pm102$ & \lyb\ 1026 & --- & --- & --- & --- & bl with \ovi\ \\
& & \lya\ 1216 & --- & --- & --- & --- & nm \\
& & \ovi\ 1032 & 4265.75 & 30 & $14.41\pm0.07$ & 1.0$^b$ & bl, adopt $b$ from \civ\ \\
& & \ovi\ 1038 & 4289.31 &--- & --- & ---& ---\\
& & \nv\ 1239 & 5121.03 & 30 & $13.89\pm0.04$ & 1.0$^b$ & bl, adopt $b$ from \civ\ \\
& & \nv\ 1243 & 5137.49 & --- & --- & ---& bl\\
& & \civ\ 1548 & 6399.95 & $29.6\pm0.6$ & $13.97\pm0.01$ & 1.0$^b$ & bl\\
& & \civ\ 1551 & 6410.60 & --- & --- & --- & ---\\
\hline
13 & 3.1344 & \lyc\ 973 & 4022.73 & 28 & $<14.9$ & 1.0$^c$ & bl, adopt $b$ from \civ\ \\
& $394\pm102$ & \lyb\ 1026 & --- & --- & --- & --- & bl with \ovi\ \\
& & \lya\ 1216 & --- & --- & --- & --- & nm \\
& & \ovi\ 1032 & 4266.37 & 28 & $14.78\pm0.07$ & $1.0^b$ & bl, adopt $b$ from \civ\ \\
& & \ovi\ 1038 & 4289.94 & --- &--- & ---& ---\\
& & \nv\ 1239 & 5121.78 & 28 & $14.00\pm0.03$ & $1.0^b$ & bl, adopt $b$ from \civ\ \\
& & \nv\ 1243 & 5138.23 &--- & ---& ---& ---\\
& & \civ\ 1548 & 6400.88 & $27.5\pm0.6$ & $13.88\pm0.01$ & 1.0$^b$ & bl\\
& & \civ\ 1551 & 6411.53 & ---& ---& ---& ---\\
\hline
14 & 3.1355 & \lyc\ 973 & --- & 21 & $<14.8$ & 1.0$^c$ & bl, adopt $b$ from \civ\ \\
& $470\pm102$ & \lyb\ 1026 & --- & --- & --- & --- & bl\\
& & \lya\ 1216 & --- & --- & --- & --- & nm \\
& & \civ\ 1548 & 6402.58 & $20.5\pm1.3$ & $13.05\pm0.02$ & 1.0$^d$ & \\
& & \civ\ 1551 & 6413.23 & --- & ---&--- & \\
\hline
15 & 3.1360 & \lyb\ 1026 & --- & --- & --- & --- & bl with \ovi\ \\
& $509\pm102$ & \lya\ 1216 & --- & --- & --- & --- & nm \\
& & \ovi\ 1032 & 4268.02 & 7 & $13.91\pm0.02$ & $1.0^b$ & adopt $b$ from \civ\ \\
& & \ovi\ 1038 & 4291.60 & --- & ---& --- & \\
& & \nv\ 1239 & 5123.76 & 7 & $13.00\pm0.10$ & $1.0^c$ & adopt $b$ from \civ\ \\
& & \nv\ 1243 & 5140.22 & --- & --- & ---& bl\\
& & \civ\ 1548 & 6403.36 & $6.6\pm0.3$ & $13.26\pm0.02$ & 1.0$^b$ & \\
& & \civ\ 1551 & 6414.01 & --- & --- & --- & \\
\hline
16 & 3.1366 & \lyb\ 1026 & --- & --- & --- & --- & bl with \ovi\ \\
& $554\pm102$ & \lya\ 1216 & --- & --- & --- & --- & nm \\
& & \ovi\ 1032 & 4268.64 & 9 & $14.16\pm0.02$ & $1.0^b$ & adopt $b$ from \civ\ \\
& & \ovi\ 1038 & 4292.22 & --- & --- & --- & \\
& & \nv\ 1239 & 5124.50 & 9 & $12.85\pm0.11$ & $1.0^c$ & adopt $b$ from \civ\ \\
& & \nv\ 1243 & 5140.97 & --- & --- & --- & bl\\
& & \civ\ 1548 & 6404.28 & $8.5\pm0.7$ & $12.79\pm0.03$ & 1.0$^d$ & \\
& & \civ\ 1551 & 6414.94 & --- &--- & ---& \\
\hline
17 & 3.1373 & \lyc\ 973 & 4025.61 & $15.0\pm1.0$ & $\leq14.45$ & 1.0$^c$ & bl?\\
& $601\pm102$ & \lyb\ 1026 & --- & --- & --- & --- & bl with \ovi\ \\
& & \lya\ 1216 & --- & --- & --- & --- & nm \\
& & \ovi\ 1032 & 4269.36 & 8 & $14.07\pm0.03$ & $1.0^b$ & bl, adopt $b$ from \civ\ \\
& & \ovi\ 1038 & 4292.95 & --- & --- & --- & ---\\
& & \nv\ 1239 & 5125.37 & 8 & $12.82\pm0.10$ & $1.0^d$ & adopt $b$ from \civ\ \\
& & \nv\ 1243 & 5141.84 & --- & --- &--- & ---\\
& & \civ\ 1548 & 6405.37 & $7.7\pm0.3$ & $13.06\pm0.01$ & 1.0$^d$ & \\
& & \civ\ 1551 & 6416.03 & --- & ---&--- &\\
\hline
18 & 3.1377 & \lyb\ 1026 & --- & --- & --- & --- & bl with \ovi\ \\
& $636\pm102$ & \lya\ 1216 & --- & --- & --- & --- & nm \\
& & \ovi\ 1032 & 4269.78 & 5 & $13.69\pm0.03$ & $1.0^b$ & bl, adopt $b$ from \civ\  \\
& & \ovi\ 1038 & 4293.36 & --- & --- & --- & ---\\
& & \nv\ 1239 & 5125.87 & 5 & $12.62\pm0.12$ & $1.0^d$ & adopt $b$ from \civ\ \\
& & \nv\ 1243 & 5142.33 & --- & --- & --- &\\
& & \civ\ 1548 & 6405.99 & $4.6\pm0.2$ & $12.95\pm0.02$ & 1.0$^d$ & \\
& & \civ\ 1551 & 6416.65 & --- & --- & --- & \\
\enddata
\tablenotetext{a}{Unsaturated doublets, where we solve for $C_0$ from fitting.}
\tablenotetext{c}{Unsaturated single lines, where we adopt $C_0=1$ because it is not constrained.}
\tablenotetext{d}{Weak doublets, where we take a conservative approach by setting $C_0=1$.}
\end{deluxetable*}

\clearpage
\startlongtable
\begin{deluxetable*}{cccccccc}
\tablecaption{Individual absorption lines of J1225+4831. See \Cref{tab:Q0249} for descriptions of the table contents. For the notes, sat=saturated line, bl=blended with neighboring systems or unrelated lines (e.g., lines in the Lyman forest), nm=no measurements.\label{tab:J1225}}
\tabletypesize{\scriptsize}
\tablehead{
\colhead{$\#$ } & \colhead{$z_{abs}$} & 
\colhead{ID} & \colhead{$\lambda_{obs}$} & 
\colhead{$b$} & \colhead{$\log N$} & 
\colhead{$C_0$ } & \colhead{Notes}\\ 
\colhead{} & \colhead{$\textrm{v}$ (\kms)} & \colhead{} & \colhead{(\AA)} &  \colhead{(\kms) } &
\colhead{(\cmN)} & \colhead{} & \colhead{} 
} 
\startdata
1 & 3.0801 & \lya\ 1216 & 4959.88 & $18.1\pm3.2$ & $12.54\pm0.09$ & 1.0$^c$ & bl\\
& $-1168\pm103$ & \civ\ 1548 & 6316.81 & $10.1\pm8.3$ & $ 12.72\pm0.25$ & 1.0$^b$ & \\
& & \civ\ 1551 & 6327.30 & --- & ---& ---&\\
\hline
2 & 3.0807 & \lya\ 1216 & 4960.77 & $70.5\pm2.5$ & $13.48\pm0.02$ & 1.0$^c$ & bl\\
& $-1121\pm103$ & \civ\ 1548 & 6317.74 & $61.9\pm18.1$ & $13.61\pm0.11$ & 1.0$^b$ & bl\\
& & \civ\ 1551 & 6328.23 & & & & \\
\hline
3 & 3.0826 & \lyb\ 1026 & --- & --- & --- & --- & bl\\
& $-980\pm103$ & \lya\ 1216 & 4963.14 & $42.7\pm2.4$ & $13.72\pm0.02$ & 1.0$^c$ & \\
& & \civ\ 1548 & 6320.68 & $18.6\pm2.7$ & $13.75\pm0.06$ & 1.0$^c$ & \\
& & \civ\ 1551 & 6331.17 & --- & ---& --- & bl\\
\hline
4 & 3.0847 & \lya\ 1216 & 4965.70 & $29.4\pm1.4$ & $12.84\pm0.02$ & 1.0$^c$ &  \\
& $-826\pm103$ & \civ\ 1548 & 6323.93 & $23.5\pm3.1$ & $12.70\pm0.05$ & 1.0$^b$ & \\
& & \civ\ 1551 & 6334.43 & --- & --- & --- & \\
\hline
5 & 3.0895 & \lyb\ 1026 & --- & --- & --- & --- & bl \\
& $-477\pm103$ & \lya\ 1216 & 4971.49 & $42.6\pm1.2$ & $13.58\pm0.02$ & 1.0$^c$ & \\
& & \nv\ 1239 & --- & --- & --- & --- & nm \\
& & \nv\ 1243 & 5082.43 & $36.9\pm2.7$ & $14.22\pm0.03$ & 1.0$^c$ &  \\
& & \siiv\ 1394 & 5699.76 & $46.1\pm6.3$ & $13.00\pm0.05$ & 1.0$^b$ &\\
& & \siiv\ 1403 & 5736.63 & --- &--- & ---& \\
& & \civ\ 1548 & 6331.36 & $37.1\pm0.7$ & $13.88\pm0.01$ & 1.0$^c$ & bl\\
& & \civ\ 1551 & 6341.87 & --- & --- & ---& ---\\
\hline
6 & 3.0941 & \lye\ 938 & 3839.47 & $44.2\pm0.9$ & $15.89^{+1.09}_{-1.25}$ & 1.0$^a$ & bl\\
& $-139\pm103$ & \lyd\ 950 & --- & --- & --- & --- & bl\\
& & \lyc\ 973 & 3981.67 & --- & --- & --- & bl, sat\\
& & \lyb\ 1026 & 4199.20 & --- & --- & --- & bl, sat\\
& & \lya\ 1216 & 4977.09 & --- & --- & --- & bl, sat\\
& & \nv\ 1239 & --- & --- & --- & --- & nm \\
& & \nv\ 1243 & 5088.15 & $55.9\pm1.9$ & $15.16\pm0.08$ & 1.0$^c$ & \\
& & \siiv\ 1394 & 5706.17 & $34.0\pm1.8$ & $13.42\pm0.02$ & 1.0$^b$ & \\
& & \siiv\ 1403 & 5743.08 & --- & --- & ---&\\
& & \civ\ 1548 & 6338.49 & 34 & $>14.7$ & 1.0$^a$ & sat, adopt $b$ from \siiv\ \\
& & \civ\ 1551 & 6349.01 & ---&--- &--- & ---\\
\hline
7 & 3.0956 & \lyb\ 1026 & --- &--- & --- & --- & bl \\
& $-32\pm103$ & \lya\ 1216 & 4978.84  & 8 & $13.62\pm0.20$ & 1.0$^c$ & bl, adopt $b$ from \civ\ \\
& & \nv\ 1239 & --- & --- & --- & --- & nm \\
& & \nv\ 1243 & 5090.01 & $8.4\pm1.9$ & $13.50\pm0.08$ & 1.0$^c$ & \\
& & \civ\ 1548 & 6340.81 & $8.0\pm0.3$ & $13.36\pm0.01$ & 1.0$^b$ & \\
& & \civ\ 1551 & 6351.33 & --- & --- & --- & \\
\hline
8 & 3.0960 & \lyb\ 1026 & --- & --- &--- & --- & bl \\
& $-2\pm103$ & \lya\ 1216 & 4979.35 & 18 & $13.81\pm0.03$ & 1.0$^c$ & bl, adopt $b$ from \civ\ \\
& & \nv\ 1239 & --- & --- & --- & --- & nm \\
& & \nv\ 1243 & 5090.51 & $17.8\pm2.1$ & $13.96\pm0.04$ & 1.0$^c$ & \\
& & \civ\ 1548 & 6341.43 & $17.5\pm0.4$ & $13.79\pm0.01$ & 1.0$^b$ & \\
& & \civ\ 1551 & 6351.95 & --- &--- &--- & \\
\enddata
\tablenotetext{a}{Heavily saturated lines, whose $C_0$ equals the observed depth of the line.}
\tablenotetext{b}{Unsaturated doublets, where we solve for $C_0$ using from fitting.}
\tablenotetext{c}{Unsaturated single lines, where we adopt $C_0=1$ because it is not constrained.}
\end{deluxetable*}

\clearpage
\startlongtable
\begin{deluxetable*}{cccccccc}
\tablecaption{Individual absorption lines of J1307+1230. See \Cref{tab:Q0249} for descriptions of the table contents. For the notes, sat=saturated line, bl=blended with neighboring systems or unrelated lines (e.g., lines in the Lyman forest).\label{tab:J1307}}
\tabletypesize{\scriptsize}
\tablehead{
\colhead{$\#$ } & \colhead{$z_{abs}$} & 
\colhead{ID} & \colhead{$\lambda_{obs}$} & 
\colhead{$b$} & \colhead{$\log N$} & 
\colhead{$C_0$ } & \colhead{Notes}\\ 
\colhead{} & \colhead{$\textrm{v}$ (\kms)} & \colhead{} & \colhead{(\AA)} &  \colhead{(\kms) } &
\colhead{(\cmN)} & \colhead{} & \colhead{} 
} 
\startdata
1 & 3.1771 & \lye\ 938 & 3917.62 & $35.1\pm1.4$ & $15.61\pm0.02$ & 1.0$^{a,b}$ &\\
& $-2371\pm221$ & \lyd\ 950 & 3967.49 & ---  & ---  & --- &  \\
& & \lyc\ 973 & 4062.71 & --- & --- & --- &  \\
& & \lyb\ 1026 & 4284.88 & --- & --- & --- & sat\\
& & \lya\ 1216 & 5078.39 & --- & --- & --- & sat \\
& & \ciii\ 977 & 4081.03 & $45.4\pm16.2$ & $13.23\pm0.12$ & 1.0$^c$ & \\
& & \civ\ 1548 & 6466.99 & $17.3\pm4.9$ & $13.14\pm0.10$ & 1.0$^b$ & \\
& & \civ\ 1551 & 6477.72 & --- & --- & --- &\\
\hline
2 & 3.1797 & \lye\ 938 & 3919.69 & $44.3\pm3.8$ & $15.69\pm0.03$ & 1.0$^{a,b}$ &  \\
& $-2186\pm221$ & \lyd\ 950 & 3969.59 & --- & --- & --- &  \\
& & \lyc\ 973 & 4064.86 & --- & --- & --- & bl \\
& & \lyb\ 1026 & 4287.14 & --- & --- & --- & sat \&\ bl\\
& & \lya\ 1216 & 5081.07 & --- & --- & --- & sat \&\ bl \\
& & \ciii\ 977 & 4083.65 & 40 & $>14.0$ & 1.0$^a$ & bl \&\ sat, adopt $b$ from \civ\ \\
& & \civ\ 1548 & 6471.01 & $40.0\pm8.2$ & $13.53\pm0.07$ & 1.0$^b$ & \\
& & \civ\ 1551 & 6481.75 & --- & ---& ---& \\
\hline
3 & 3.1807 & \lye\ 938 & 3920.74 & $23.1\pm2.2$ & $15.28\pm0.04$ & 1.0$^{a,b}$ &  \\
& $-2116\pm221$ & \lyd\ 950 & 3970.66 & --- & --- & --- &  \\
& & \lyc\ 973 & 4065.95 &---  & --- & --- & bl \\
& & \lyb\ 1026 & 4288.30 & --- & --- & --- & sat \&\ bl\\
& & \lya\ 1216 & 5082.44 & --- & --- & --- & sat \&\ bl \\
& & \ciii\ 977 & 4084.63 & 18 & $>13.7$ & 1.0$^a$ & bl \&\ sat, adopt $b$ from \civ\ \\
& & \siiv\ 1394 & 5826.87 & $29.7\pm4.1$ & $12.83\pm0.07$ & 1.0$^b$ &\\
& & \siiv\ 1403 & 5864.56 & --- & --- & --- &\\
& & \civ\ 1548 & 6472.56 & $18.1\pm3.5$ & $13.39\pm0.07$ & 1.0$^b$ & \\
& & \civ\ 1551 & 6483.30 & --- & --- & --- & \\
\hline
4 & 3.2030 & \lye\ 938 & 3941.53 & --- & --- & --- & bl\\
& $-526\pm221$ & \lyd\ 950 & 3991.71 & $27.5\pm1.1$ & $15.33\pm0.03$ & 1.0$^{a,b}$ & \\
& & \lyc\ 973 & 4087.51 & --- & --- & --- &  \\
& & \lyb\ 1026 & 4311.04 & --- & --- & --- & sat\\
& & \lya\ 1216 & 5109.39 & --- & --- & --- & sat \&\ bl \\
& & \ciii\ 977 & 4106.42 & $15.6\pm5.1$ & $13.61\pm0.08$ & 1.0$^c$ & \\
& & \siiv\ 1394 & 5857.95 & $11.1\pm3.2$ & $12.58\pm0.10$ & 1.0$^d$ & \\
& & \siiv\ 1403 & 5895.84 & --- & --- & ---&\\
& & \civ\ 1548 & 6507.08 & $20.8\pm3.1$ & $13.56\pm0.06$ & 1.0$^b$ & \\
& & \civ\ 1551 & 6517.89 & --- &--- & ---& \\
\hline
5 & 3.2039 & \lye\ 938 & 3942.43 & 18 & $14.73\pm0.22$ & 1.0$^{a,b}$ & bl, adopt $b$ from \civ\ \\
& $-464\pm221$ & \lyd\ 950 & 3992.62 & --- & --- & --- & bl\\
& & \lyc\ 973 & 4088.44 & --- & --- & --- & bl \\
& & \lyb\ 1026 & 4312.02 & --- & --- &---  & sat \&\ bl\\
& & \lya\ 1216 & 5110.55 & --- & --- & --- & sat \&\ bl \\
& & \ciii\ 977 & 4107.29 & $11.6\pm4.4$ & $13.49\pm0.20$ & 1.0$^c$ & \\
& & \siiv\ 1394 & 5859.21 & $15.4\pm2.5$ & $12.73\pm0.06$ & 1.0$^d$ & \\
& & \siiv\ 1403 & 5897.10 & --- & --- & ---&\\
& & \civ\ 1548 & 6508.48 & $17.8\pm2.1$ & $13.79\pm0.04$ & 1.0$^b$ & \\
& & \civ\ 1551 & 6519.28 & --- & --- & --- & \\
\hline
6 & 3.2047 & \lye\ 938 & 3943.17 & 21 & $15.89\pm0.15$ & 1.0$^{a,b}$ & bl, adopt $b$ from \civ\ \\
& $-407\pm221$ & \lyd\ 950 & 3993.38 & --- & --- & --- & bl\\
& & \lyc\ 973 & 4089.22 & --- &---  & --- & sat \&\ bl \\
& & \lyb\ 1026 & 4312.83 & --- & --- & --- & sat \&\ bl\\
& & \lya\ 1216 & 5111.52 & --- & --- & --- & sat \&\ bl \\
& & \ciii\ 977 & 4108.08 & 21 & $14.17\pm0.18$ & 1.0$^c$ & bl, adopt $b$ from \civ\ \\
& & \siiv\ 1394 & 5860.32 & $16.3\pm1.2$ & $13.13\pm0.03$ & 1.0$^b$ & bl\\
& & \siiv\ 1403 & 5898.23 & --- & --- & ---& ---\\
& & \civ\ 1548 & 6509.72 & $20.5\pm2.5$ & $14.02\pm0.04$ & 1.0$^b$ & \\
& & \civ\ 1551 & 6520.52 & --- & --- & --- & \\
\hline
7 & 3.2051 & \lye\ 938 & 3943.60 & 15 & $15.39\pm0.30$ & 1.0$^{a,b}$ & bl, adopt $b$ from \civ\ \\
& $-375\pm221$ & \lyd\ 950 & 3993.80 & --- & --- & --- & bl\\
& & \lyc\ 973 & 4089.66 & --- & --- &---  & sat \&\ bl \\
& & \lyb\ 1026 & 4313.30 & --- & --- & --- & sat \&\ bl\\
& & \lya\ 1216 & 5112.06 & --- & --- & --- & sat \&\ bl \\
& & \ciii\ 977 & 4108.47 & 15 & $14.75\pm0.22$ & 1.0$^c$ & bl, adopt $b$ from \civ\ \\
& & \siiv\ 1394 & 5860.88 & $9.5\pm1.4$ & $12.63\pm0.06$ & 1.0$^d$ & bl\\
& & \siiv\ 1403 & 5898.79 & --- & --- & --- & ---\\
& & \civ\ 1548 & 6510.34 & $15.0\pm2.0$ & $13.72\pm0.06$ & 1.0$^b$ & \\
& & \civ\ 1551 & 6521.14 & --- & --- & --- & \\
\enddata
\tablenotetext{a}{Heavily saturated lines, whose $C_0$ equals the observed depth of the line.}
\tablenotetext{b}{Unsaturated doublets, where we solve for $C_0$ from fitting.}
\tablenotetext{c}{Unsaturated single lines, where we adopt $C_0=1$ because it is not constrained.}
\tablenotetext{d}{Weak doublets, where we take a conservative approach by setting $C_0=1$.}
\end{deluxetable*}

\clearpage
\startlongtable
\begin{deluxetable*}{cccccccc}
\tablecaption{Individual absorption lines of J1633+1411. See \Cref{tab:Q0249} for descriptions of the table contents. For the notes, bl=blended with neighboring systems or unrelated lines (e.g., lines in the Lyman forest), nm=no measurements.\label{tab:J1633}}
\tabletypesize{\scriptsize}
\tablehead{
\colhead{$\#$ } & \colhead{$z_{abs}$} & 
\colhead{ID} & \colhead{$\lambda_{obs}$} & 
\colhead{$b$} & \colhead{$\log N$} & 
\colhead{$C_0$ } & \colhead{Notes}\\ 
\colhead{} & \colhead{$\textrm{v}$ (\kms)} & \colhead{} & \colhead{(\AA)} &  \colhead{(\kms) } &
\colhead{(\cmN)} & \colhead{} & \colhead{} 
} 
\startdata
1 & 4.2279 & \nv\ 1239 & --- & --- & --- & --- & bl \\
& $-8224\pm100$ & \nv\ 1243 & --- & --- & --- & --- & bl\\
& & \civ\ 1548 & 8093.83 & $29.9\pm2.6$ & $14.28\pm0.11$ & $0.44\pm0.05^b$ & \\
& & \civ\ 1551 & 8107.27 &--- & ---& ---& \\
\hline
2 & 4.2347 & \nv\ 1239 & --- & --- & --- & --- & bl \\
& $-7848\pm100$ & \nv\ 1243 & --- & --- & --- & --- & bl\\
&& \civ\ 1548 & 8104.36 & $52.7\pm9.5$ & $14.18\pm0.18$ & $0.49\pm0.17^b$ &\\
& & \civ\ 1551 & 8117.82 & --- & --- & --- &\\
\hline
3 & 4.2380 & \nv\ 1239 & --- & --- & --- & --- & bl \\
& $-7660\pm100$ & \nv\ 1243 & --- & --- & --- & --- & bl\\
& & \civ\ 1548 & 8109.55 & $35.6\pm9.2$ & $13.43\pm0.66$ & $0.61\pm0.19^b$ & bl \\
& & \civ\ 1551 & 8123.01 & --- & --- & --- & ---\\
\hline
4 & 4.2512 & \nv\ 1239 & 6505.30 & $44.4\pm0.82$ & $14.19\pm0.03$ & $0.58\pm0.02^b$ &  \\
& $-6923\pm100$ & \nv\ 1243 & 6526.21 & --- & --- & --- & bl\\
& & \civ\ 1548 & 8129.97 & $22.8\pm2.0$ & $14.65\pm0.12$ & $0.55\pm0.03^b$ &  \\ 
& & \civ\ 1551 & 8143.47 & --- & ---& ---&\\
\hline
5 & 4.2580 & \nv\ 1239 & --- & --- & --- & --- & bl \\
& $-6550\pm100$ & \nv\ 1243 & --- & --- & --- & --- & \\
& & \civ\ 1548 & 8140.33 & $25.2\pm1.9$ & $14.13\pm0.11$ & $0.80\pm0.11^b$ &  \\ 
& & \civ\ 1551 & 8153.84 & --- & --- & --- &\\
\hline
6 & 4.2622 & \nv\ 1239 & 6518.92 & 51 & $<14.55$ & $\geq0.2^b$ & bl, adopt $b$ from \civ\ \\
& $-6314\pm100$ & \nv\ 1243 & 6539.88 & --- & --- & --- & ---\\
& & \civ\ 1548 & 8146.88 & $50.7\pm3.3$ & $14.50\pm0.02$ & $0.60\pm0.02^b$ & \\ 
& & \civ\ 1551 & 8160.41 & ---& ---&--- & \\
\hline
7 & 4.2644 & \nv\ 1239 & 6518.92 & 35 & $<13.58$ & $\geq0.2^b$ & bl, adopt $b$ from \civ\ \\
& $-6188\pm100$ & \nv\ 1243 & 6521.65 & --- & --- & --- & ---\\
& & \civ\ 1548 & 8150.37 & $35.1\pm34.5$ & $<14.6$ & $\geq0.4^b$ & bl \\ 
& & \civ\ 1551 & 8163.90 & --- & --- & --- & ---\\
\hline
8 & 4.2811 & \nv\ 1239 & 6542.34 & $62.3\pm5.5$ & $<15.6$ & $\geq0.2^b$ & bl \\
& $-5255\pm100$ & \nv\ 1243 & 6563.37 & --- & --- & --- & \\
& & \civ\ 1548 & 8176.27 & $51.6\pm4.1$ & $<15.1$ & $\geq0.2^b$ & bl \\ 
& & \civ\ 1551 & 8189.84 & --- & ---& ---& nm for part of the line\\
\hline
9 & 4.2843 & \nv\ 1239 & 6546.30 & $57.5\pm1.6$ & $14.86\pm0.03$ & $0.33\pm0.01^b$ &  \\
& $-5072\pm100$ & \nv\ 1243 & 6567.35 & --- & --- & --- & \\
& & \civ\ 1548 & 8181.21 & $98.5\pm10.8$ & $\geq14.24$ & 1.0$^c$ &  \\ 
& & \civ\ 1551 & --- & --- & ---& ---& nm\\
\hline
10 & 4.3675 & \nv\ 1239 & 6649.37 & $12.3\pm0.7$ & $13.02\pm0.29$ & 1.0$^d$ &  \\
& $-435\pm100$ & \nv\ 1243 & 6670.75 & --- & --- & --- & \\
& & \civ\ 1548 & 8309.96 & $15.8\pm1.7$ & $13.49\pm0.25$ & 1.0$^b$ &  \\ 
& & \civ\ 1551 & 8323.75 & --- & ---& ---& \\
\enddata
\tablenotetext{b}{Unsaturated doublets, where we solve for $C_0$ from fitting.}
\tablenotetext{c}{Unsaturated single lines, where we adopt $C_0=1$ because it is not constrained.}
\tablenotetext{d}{Weak doublets, where we take a conservative approach by setting $C_0=1$.}
\end{deluxetable*}

\clearpage
\startlongtable
\begin{deluxetable*}{ccccccccc}
\tablecaption{Parameters of some absorbers. Columns show quasar name, component, velocity shifts in \kms, ionisation parameter, total hydrogen column density, and C, Si, N, O abundances.\label{tab:parameter}}
\tablehead{
\colhead{QSO} & \colhead{Component} & \colhead{Velocity} & \colhead{$\log U$} & \colhead{$\log N(\mbox{H})$} &  \colhead{[C/H]} & \colhead{[Si/H]} & \colhead{[N/H]} & \colhead{[O/H]}\\
\colhead{}  & \colhead{} & \colhead{(\kms)} & \colhead{} & \colhead{(\cmN)} & \colhead{} & \colhead{} &  \colhead{} & \colhead{}
}  
\startdata
Q0249$-$222   & 3 & $-7424$ & $-0.7\pm0.2$ & $19.0\pm0.3$ & $-0.9\pm0.2$ & $-0.8\pm0.3$ & --- & ---\\ 
  & 4 & $-7373$ & $-1.3\pm0.1$ & $18.8\pm0.2$ & $-1.2\pm0.1$ & $-1.3\pm0.1$ & --- & ---\\
  & 6 & $-7138$ & $-0.5\pm0.1$ & $19.2\pm0.1$ & $-1.0\pm0.1$ & $-0.9\pm0.1$ & --- & ---\\
  & 7 & $-7066$ & $-1.6\pm0.1$ & $18.0\pm0.1$ & $-0.9\pm0.2$ & $-0.9\pm0.1$ & --- & ---\\   
\hline
J1008+3623 & 1 & $-3108$ & $-0.2\pm0.1$ & $18.6\pm0.2$ & $-0.1\pm0.1$ & --- & $0.0\pm0.1$ & $-0.1\pm0.2$\\ 
 & 2 & $-2942$ & $0.6\pm0.2$ & $19.7\pm0.3$ & $0.3\pm0.3$ & --- & $0.1\pm0.3$ & ---\\ 
 & 3 & $-2807$ & $-0.6\pm0.1$ & $18.7\pm0.2$ & $0.1\pm0.1$ & $0.1\pm0.2$ & $0.4\pm0.3$ & --- \\ 
 & 4 & $-2712$ & $-0.5\pm0.2$ & $<18.7$ & $>0.2$ & --- & $>0.1$ & ---\\
 & 6 & $-2503$ & $-0.2\pm0.1$ & $<19.5$ & $>-0.3$ & $>0.1$ & $>-0.3$ & $>-0.4$\\
 & 8 & $-1238$ & $-0.5\pm0.2$ & --- & --- & --- & --- & ---\\
 & 9 & $-243$ & $-0.3\pm0.3$ & $20.1\pm0.4$ & $-1.2\pm0.1$ & $-1.0\pm0.6$ & $-0.8\pm0.4$ & $-1.4\pm0.6$ \\
 & 10 & $-210$ & $-0.3\pm0.1$ & $19.5\pm0.6$ & $-0.4\pm0.2$ & --- & $-0.7\pm0.4$ & $-0.8\pm0.8$\\
 & 11 & $-102$ & $-0.1\pm0.2$ & $19.8\pm0.3$ & $-1.1\pm0.1$ & --- & $-0.9\pm0.2$ & $-1.1\pm0.3$ \\
 & 12 & $346$ & $-0.5\pm0.1$ & $<19.8$ & $>-1.2$ & --- & $>-1.5$ & $>-1.5$\\
 & 13 & $394$ & $-0.2\pm0.1$ & $<20.2$ & $>-1.3$ & --- & $>-1.5$ & $>-1.7$\\
 & 15 & $509$ & $-0.4\pm0.2$ & --- & --- & --- & --- & ---\\
 & 16 & $554$ & $-0.3\pm0.2$ & --- & --- & --- & --- & ---\\
 & 17 & $601$ & $-0.3\pm0.2$ & $\leq19.8$ & $\geq -1.7$ & --- & $\geq -2.5$ & $\geq -1.9$\\
 & 18 & $636$ & $-0.3\pm0.2$ & --- & --- & --- & --- & ---\\
\hline
J1225+4831 & 5 & $-477$ & $-0.6\pm0.3$ & $18.3\pm0.3$ & $0.0\pm0.1$ & $1.0\pm0.3$ & $0.5\pm0.4$ & ---\\ 
 & 6 & $-139$ & $-0.6\pm0.1$ & $19.2$ to $21.8$ & $>-2.6$ & $-2.2$ to $0.6$ & $-2.2$ to $0.5$ & ---\\  
 & 7 & $-32$ & $-0.4\pm0.1$ & $18.6\pm0.4$ & $-0.5\pm0.2$ & --- & $-0.5\pm0.3$ & ---\\
 & 8 & $-2$ & $-0.4\pm0.1$ & $18.8\pm0.1$ & $-0.3\pm0.1$ & --- & $-0.3\pm0.2$ & ---\\
\hline
J1307+1230 & 1 & $-2371$ & $-1.8\pm0.2$ & $19.0\pm0.2$ & $-1.8\pm0.2$ & --- & --- & ---\\
  & 4 & $-526$ & $-1.6\pm0.1$ & $18.9\pm0.1$ & $-1.5\pm0.2$ & $-1.4\pm0.2$ & --- & ---\\ 
  & 5 & $-464$ & $-1.5\pm0.1$ & $18.4\pm0.4$ & $-0.8\pm0.3$ & $-0.7\pm0.3$ & --- & ---\\ 
  & 6 & $-407$ & $-1.8\pm0.2$ & $19.3\pm0.3$ & $-1.1\pm0.5$ & $-1.2\pm0.4$ & --- & ---\\
  & 7 & $-375$ & $-1.5\pm0.1$ & $19.1\pm0.4$ & $-1.3\pm0.6$ & $-1.4\pm0.4$ & --- & ---\\  
\hline
J1633+1411 & 4 & $-6923$ & $-0.9\pm0.2$ & --- & --- & --- & --- & ---\\
 & 10 & $-435$ & $-0.8\pm0.4$ & --- & --- & --- & --- & ---\\
\enddata
\end{deluxetable*}
\clearpage

\section*{acknowledgments}

We are grateful to Serena Perrotta for valuable discussion on the interpretation of quasar associated absorption lines. We thank the anonymous referee for useful comments and suggestions. This work was supported by funds from University of California, Riverside, USA; Sun Yat-Sen University, China, by grant AST-1009628 from the USA National Science Foundation, and by NFSC grant U1931102. 

\bibliographystyle{aasjournal}
\bibliography{reference}

\end{document}